# Impact of $\eta_{earth}$ on the capabilities of affordable space missions to detect biosignatures on extrasolar planets

Revised version v4.5 (2015/04/17)

<u>Short title</u>: Impact of $\eta_{earth}$…

Alain Léger[1,2], Denis Defrère[3], Fabien Malbet[4], Lucas Labadie[5], Olivier Absil[6,7]

[1] IAS, Univ. Paris-Sud, Orsay, France

[2] IAS, CNRS (UMR 8617), bât 121, Univ. Paris-Sud, F-91405 Orsay, France

[3] Steward Observatory, Department of Astronomy, University of Arizona, 933 N. Cherry Ave, Tucson, AZ 85721

[4] UJF-Grenoble 1 / CNRS-INSU, Institut de Planétologie et d'Astrophysique de Grenoble (IPAG), UMR 5274, BP 53, F-38041 Grenoble cedex 9, France;

[5] I. Physikalisches Institut der Universität zu Köln, Zülpicher Str. 77 50937 Cologne - Germany

[6] Département d'Astrophysique, Géophysique & Océanographie, Université de Liège, 17 Allée du Six Août, B-4000 Liège, Belgium

[7] F.R.S.-FNRS Research Associate

Corresponding author: Alain.Leger@ias.u-psud.fr

## Abstract

We present an analytic model to estimate the capabilities of space missions dedicated to the search for biosignatures in the atmosphere of rocky planets located in the habitable zone of nearby stars. Relations between performance and mission parameters such as mirror diameter, distance to targets, and radius of planets, are obtained. Two types of instruments are considered: coronagraphs observing in the visible, and nulling interferometers in the thermal infrared. Missions considered are: single-pupil coronagraphs with a 2.4 m primary mirror, and formation flying interferometers with 4 x 0.75 m collecting mirrors. The numbers of accessible planets are calculated as a function of $\eta_{earth}$. When *Kepler* gives its final estimation for $\eta_{earth}$, the model will permit a precise assessment of the potential of each instrument. Based on current estimations, $\eta_{earth}$ = 10% around FGK stars and 50% around M stars, *the coronagraph could study in spectroscopy only ~ 1.5 relevant planets, and the interferometer ~ 14.0.* These numbers are obtained under *the* major hypothesis that the *exozodiacal light* around the target stars is low enough for each instrument. In both cases, a prior detection of planets is assumed and a target list established. For the long-term future, building both types of spectroscopic instruments, and using them on the same targets, will be the optimal solution because they provide complementary information. But as a first affordable space mission, *the interferometer looks the more promising in term of biosignature harvest.*

**Key words:** astrobiology, exoplanets, coronagraph, starshade, nulling interferometers

# 1. Introduction

Exobiology outside the Solar System is a new field of science that should become accessible in the mid-term future if suitable instruments can be funded and built. They should be able to give us the first scientific answers to questions that humanity has been asking itself for over 2000 yrs (Democritus of Abdera 460-371 BC, Epicurus of Samos 341-270 BC).

In a founding article, Tobias Owen (1980) states: "… it is not a simple accident that life on Earth is based on carbon with water as its liquid. It seems possible that most of life elsewhere in the universe will rely on this same chemistry (…). With our present understanding of the requirements for life (…) we find some real support for carbon-water chauvinism". This statement has an important implication, the existence of bio-signatures that can be searched for by remote sensing. The real way for a life based on carbon chemistry, having $CO_2$ (fully oxidized C) as an abundant raw material, is to synthesize its organic carbon (partially reduced C) from it, using the most abundant source of free energy on a planet, the stellar radiation. A frequent output of this reduction of $CO_2$ is $O_2$, which can be search for in the exoplanetary atmospheres. This biosignature has possible false negatives, but seems having no false positive, because up to now (2015) the many attempts to falsify it have failed. Making the case of our planet, Owen pointed: "The huge abundance of free oxygen is very difficult to explain without invoking biology, since this highly reactive gas would rapidly combine with the crust". He also noted: "the A-band of $O_2$ at 7600 Å is remarkably strong". The way to go was traced.

Later, Angel et al. (1986) first -- followed by Léger et al. (1993), Selsis et al. (2002), and other authors -- showed that a solid alternative to the search for $O_2$ at visible wavelengths (0.76 μm) was to look for $O_3$ (9.6 μm) in the thermal infrared (IR).

A number of questions to be discussed were put forward by several authors (Burke et al. 1992a, 1992b; Schneider 1995; Borucki et al. 1996; Beichman et al. 1996; Léger 2001; Mennesson et al. 2005), which answers would require to establish an ambitious instrumentation roadmap for the search of possible biosignatures. These questions are:

(1) Are there giant planets around other stars than the Sun?

(2) Are there telluric planets around other stars than the Sun?

(3) Are there telluric planets located in the habitable zone (HZ) of their stars, and how many?

(4) Can we identify the relevant planets around the nearest stars?

(5) Can we build a mission that can search for biosignature(s) in the latter group?

Today, we have positive answers to questions (1) and (2) (Mayor & Queloz 1996, Léger, Rouan, Schneider et al. 2009). *Kepler* was built to answer question (3) (Borucki et al. 1996), but presently there is no funded plan to answer question (4). To address goal (5), the choice of the mission to build was, for various reasons, decided before we knew the answer to question 3. The goal of the present paper is to revisit that point, and discuss how the answer to (3) – which we should get from *Kepler* – would influence our choices for answering question (5).

There are two avenues to do that: spectroscopic transits (with the JWST or one of the future extremely large ground-based telescopes), and spatially resolved spectroscopy. Only the second approach is discussed here for the sake of conciseness. Some preliminary estimates with Kepler of the $\eta_{earth}$ parameter, i.e., the mean number of Earth analogues and super-Earths in the HZ of stars, are of the order of 5% to 10% around solar type stars (Petigura et al. 2013, Batalha 2014), and 50% around M stars (Kopparapu 2013). Considering the large number of stars in our galaxy (~ $3 \times 10^{11}$), this field of science benefits from a huge reservoir of targets to be studied in the next decades, possibly centuries.

In the last decade, during the "golden age" of NGST, *Darwin*, and TPF, very large instruments were considered to conduct the spectroscopy of Earths analogs (Angel et al. 1986, Beichman et al. 1999, Cockell 2009, Defrère et al. 2010). These included for instance coronagraphs on an 8x3 m telescope (Levine et al. 2009) or on a 5 m diameter (Φ) telescope (Lunine et al. 2009), and interferometers made up of four telescopes with 2 m collecting mirrors.

In post-JWST era only missions which cost no more than ~ 2 G$ can realistically be considered by any space agency. The sizes of the corresponding instruments must be therefore significantly smaller, typically a 2.4 m mirror for the single-dish coronagraph approach, or four 0.75 m telescopes for the interferometric approach as recently proposed to ESA for the L 3 mission slot (Quirrenbach et al. 2014).

In this paper, an analytic model is described (Sect. 2 and 3) that can estimate the science return of such size-reduced instruments in terms of exo-Earths spectroscopy as a function of $\eta_{earth}$, for both visible coronagraphs (Sect.4, 5, and 7) and IR interferometers (Sect. 6, and 7). Conclusions are given in Sect.9.

Note that in most parts of this study assume that a *precursor mission has identified the list of target stars* having actually at least one telluric planet in their HZ, before the spectroscopic mission is launched.

This prior detection would save a significant fraction of time for the spectroscopic missions, thereby increasing their scientific return (see Sect. 4.7 and 6.7). There is a second, and even more important, reason for such a precursor mission. The question of the presence of extrasolar life will have a large importance in the science of our century with an impact on the general public, the tax-payer. It is our responsibility to make a statement only if we have firm indices. The example of too rapid claims by the Viking missions is still present in many minds. "Extraordinary claims require extraordinary proofs/evidence" said Marcello Truzzi (1978)/Carl Sagan (1980) in a modern version of the Laplace (1749-1827) statement, "The weight of evidence for an extraordinary claim must be proportioned to its strangeness". Detecting a gas mixture, possibly indicative of a biosignature, would be a real indication of extrasolar life only if we have a solid idea of the physical conditions at the surface of the planet, which requires *the knowledge of the planetary mass*. The mass, combined with a radius estimated from the direct imaging mission, gives the mean density of the planet and allows a first distinction between rocky planets, water ocean-planets, and planets with a hydrogen rich atmospheres. This is needed before deriving any statement on the possible presence of life on a planet.

The prior detection of Earth-like planets amenable to characterization by a high contrast imaging/interferometry mission could be done by indirect detection techniques, such as high-precision ground-based radial velocity surveys (provided that stellar convection noise can be mitigated, e.g., Meunier et al. 2010), or high-precision astrometry, which would need a dedicated space mission (e.g., Malbet et al. 2012).

However, the case of detection by the spectrometric mission itself is considered in Sect.4.7 and 6.7, with the associated disadvantages.

In this paper, an analytic model is described (Sect. 2 and 3) that can estimate the science return of such size-reduced instruments in terms of exo-Earths spectroscopy as a function of $\eta_{earth}$, for both visible coronagraphs (Sect. 4, 5, and 7) and IR interferometers (Sect. 6, and 7). Conclusions are given in Sect. 9, after a short discussion of ground-based instruments in Sect. 8.

## 2. Hypotheses for a simple model

To build a simple model, the following hypotheses are made. Systematic noises are assumed to be under control and the limiting noise is the quantum ("photon") noise, a critical hypothesis.

For the sake of simplicity, quantities such as integration times are computed only for some values of different parameters: position within the HZ around the star, observing wavelengths, and planetary radius. Explicitly:

- *Position in the HZ:* the inner and outer limits of the HZ are the object of a great activity since the pioneering paper by Kasting et al. (1993). Many recent works in astronomy, e.g., Kopparapu et al. (2013), rely on 1D, therefore cloud-free, models. For a Sun-like star, the latter find mean semi-major axis $a_{IHZ}$ = 0.99 AU for the inner boundary, and $a_{OHZ}$ = 1.70 AU for the outer. Regarding the inner boundary, a recent work in geophysics (Leconte et al. 2014) reports a detailed 3D simulation of the Earth case that includes water clouds and their feedback on the surface temperature. The authors find that if present Earth was moved from 1.0 AU to 0.96 AU, increasing the insulation by 8%, from 341 $Wm^{-2}$ to 368 $Wm^{-2}$, it would trigger a greenhouse effect leading to the vaporization of all the terrestrial oceans in a short geological time. This result supports the idea that Earth is very close to the inner limit of the solar HZ. Recently, a 3D model was used for calculating inner and outer HZ boundaries taking into account the cloud impacts (Kopparapu et al. 2014). These boundaries are found to depend on the stellar luminosity ($L$), but also on its effective temperature, the possible phase locking of the planet (around M and late K stars), and the planetary mass, pointing out that the idea of a HZ depending only on the stellar luminosity is a convenient but simplified concept.

The HZ concept is subject to different criticisms. However, it is clear that a planet *outside* this zone cannot support life based on carbon and on water as a solvent.

In the present paper, the values by Kopparapu et al. (2013) are adopted for both solar-type and M stars. Their geometrical mean value, $a_{HZ} = (a_{IHZ} * a_{OHZ})^{1/2} = (0.99 * 1.70)^{1/2} L_i^{1/2} = 1.30 \ L_i^{1/2}$ AU, is used, where $L_i$ is the stellar luminosity in solar unit. This mean value corresponds to a uniform distribution of orbits in $\log(P)$, or $\log(a)$, where $P$ is the planet orbital period, as **found** by Batalha et al. (2012).

- *Wavelengths (λ)*: in the visible, the most demanding wavelength for angular resolution is the longest one. It is necessary to permit the access to the whole domain required for the characterization of $O_2$ and $H_2O$ (0.6 - 0.8 µm, Des Marais et al. 2002). $CO_2$ requires observations up to 1.1 µm for large abundances, and up to 2.1 µm for abundances similar to terrestrial ones (DesMarais et al. 2002). This is considered as too demanding and not achievable in the mid-future context. $CO_2$ will therefore be discarded for coronagraphs, and we will use 0.8 $µm$ as the observing wavelength in this study.

 In the thermal IR, $λ$ = 10 µm is used as a proxy for the (6 - 18 µm) domain, permitting the characterization of ($H_2O$), $O_3$, and $CO_2$ (DesMarais et al. 2002). When spatial resolution is essential, the most demanding value, 18 µm, is used (Sect.6.1 et 6.3).

- *Planetary radius*: for the search of extrasolar life, the relevant planets are rocky (Owen 1980), and possibly water ocean-planets (Léger et al. 2004, Selsis et al. 2007), all without a major hydrogen content in their atmosphere.  They can be Earth-analogues ($R_{pl}$ < 1.25, in Earth radius unit), or super-Earths ($R_{pl}$ = 1.25 - 2.0, Borucki et al. 2011). When this parameter is essential, radii of 1.0, 1.5, and 2.0 are considered individually, otherwise the mean value $R_{pl}$ = 1.5 is used[1].

 Analyzing *Kepler* data, Rogers (2015) found that most rocky planets have a radius $R_{pl}$ < 1.6.  For a same mass, water ocean-planets have a lower density and may be a major component of hydrogen free planets in the range $R_{pl}$ = 1.5 − 2.0. However, their ability to harbor life is possibly lower than that of rocky planets (Tian and Ida, 2015), and the case $R_{pl}$ = 2.0 is presented in the figures with dashed lines instead of full lines, to remind that they have an interest different from that of $R_{pl}$ = 1.0-1.5 planets.

- *Parameters* such as the planet albedo, instrument yield, etc, are those adopted by the detailed studies used to adjust the model.

- *Exozodiacal light level*: its impact is discussed in Sect. 4.5 for coronagraph, and Sect. 6.5 for interferometers.  In the following, exozodiacal light is

---

[1] For direct detection spectroscopy, the signal is $α R_{pl}^2$, and integration time $α R_{pl}^4$.  When time is the limiting parameter, the mean planetary radius between 1.0 and 2.0 is 1.58.  For a radius distribution constant in log($R_{pl}$), the mean value is 1.41.  The adopted mean value of 1.50 is close to these two values.

*arbitrarily assumed* to be sufficiently low around target stars, so that it is not a major source of noise or confusion, neither for coronagraphs nor for interferometers.

In summary, the estimates of the present paper are made under favorable hypotheses, namely: (i) systematic noises have been reduced down about to the level of quantum noise, which will require major technical efforts, (ii) Exo-zodiacal light is low enough not to be a major source of noise, nor confusion. This last hypothesis must be revisited when we get real information on its level.

# 3. Required integration times

## 3.1 Signal for coronagraphs and interferometers

For an observation (*i*) the signal is proportional to the number of photo-electrons on the detector, from the planet:

$$S_i = a \, \Phi^2 \, \Delta\lambda \, R_{pl,i}^2 \, D_i^{-2} \, t_i$$

(Eq.1)

where *a* is a constant, in time$^{-1}$ unit, depending on the instrument throughput, planetary albedo…; Φ is the telescope diameter(s); $\Delta\lambda$, the spectral bandwidth; $R_{pl,i}$, the planetary radius; $D_i$, the distance of the system to us; and $t_i$, the integration time allocated to the observation.

## 3.2 Noise for coronagraphs

For a coronagraph, the main sources of noise are the quantum noise from the stellar leakage and speckles, if the latter are stable enough with time, plus the noise of exozodiacal light. Assuming the latter to be low enough (Sect. 4.5), the noise is the square root of the number of photo-electrons collected within the PSF of the planet during the integration time $t_i$ (Guyon et al. 2006, Fig.14):

$$N_i = \left[ b_{cor} \Phi^2 \, \Delta\lambda \, L_i \, \rho_i \, D_i^{-2} \, t_i \right]^{1/2}$$

(Eq.2)

where $b_{cor}$ is a constant, in time$^{-1}$ unit, analogous to *a*; $L_i$, the stellar luminosity; $\rho_i$, the residual transmission of the stellar light at the planet angular separation, a steep function of the Inner Working Angle (IWA) of the coronagraph.

## 3.3 Noise for interferometers

If instability noise (Chazelas et al. 2006, Lay 2006) can be mitigated, the main source of noise is the quantum noise from the sum of the solar zodiacal flux, the exozodiacal flux, and the stellar leakage (Defrère et al. 2010).

*Solar zodiacal flux.* When a nulling interferometer uses a single-mode fiber, the field-of-view is limited by the PSF of each collecting telescope. For an IR albedo $A$, an optical depth of the zodiacal cloud $\tau$, and a temperature of the zodiacal dust at the spacecraft position $T_{zcl}$, the solar zodiacal (SZ) flux is proportional to the surface brightness $B(\lambda, T_{zcl})$ of the zodiacal cloud multiplied by $\tau$, the solid angle of the telescope PSF, and the mirror area:

$$F_{SZ} \propto (1-A) \ B(\lambda, T_{zcl}) \tau \ (\lambda / \Phi)^2 \ \pi \ \Phi^2 \ \Delta\lambda \qquad\qquad (Eq.3)$$

Noticeably, $F_{SZ}$ is *independent of the telescope diameter $\Phi$*. Its value relative to the other noise sources such as stellar leakage is larger for smaller values of $\Phi$. It is dominating for telescopes with a diameter less than 1 m (Defrère et al. 2010).

*Exozodiacal flux.* It is proportional to $\Phi^2$ because the PSF of individual telescopes (2 arc-second in diameter at 10 µm for $\Phi$ = 1 m) is expected to contain most of the exozodiacal emission, which has a characteristic size of 0.1 as (1 AU at 10 pc). If the exozodiacal cloud was mainly symmetric, it would be removed by the detection process, and therefore produce quantum noise only (end of Sect.2). Its relative contribution to noise decreases as the telescope diameter decreases. For $\Phi$ < 1 m, and exozodiacal cloud density less than 10 times that of the SZ cloud ("10 zodi" hereafter), its is negligible compared to the contribution of the SZ cloud (Defrère et al. 2010, Sect 4 therein).

*Stellar Leakage.* A key parameter is the planetary flux / stellar leakage flux ratio. The IR flux from a planet in the HZ of its star is independent of the stellar properties (an emitting body at ~ 300 K). The IR flux from the star is proportional to its bolometric flux, $\propto L_i/D_i^2$, with a correction specific to each spectral type[2], $(T_* /T_o)^{-3}$, where $T_*$ is the stellar effective temperature and $T_o$ the solar one. This correction is important for M stars. Their bolometric contrast with a 300 K planet is more favorable than for Sun-like stars, but this correction factor decreases in the thermal IR, due to their relatively low temperatures. For a M2 star, the bolometric contrast with a 300 K planet is 30 times lower than for a G2 star, but after this correction it is only 5 times lower.

The transmission of a nulling interferometer, $\rho$, is limited by the instrument rejection ratio and the geometrical losses due to a partial resolution of the stellar disc. For F, G, and K stars (FGK thereafter), it is typically $\rho$ = 10$^{-5}$ (Defrère et al. 2010).

*Total noise* is:

$$N_i = \left[ b_{int\_1} \left( 1 + b_{int\_2} \rho \ \Phi^2 L_i \ (T_* / T_\odot)^{-3} \ D_i^{-2} \right) \Delta\lambda \ t_i \right]^{1/2} \qquad\qquad (Eq.4)$$

where the two terms in the sum correspond to the SZ cloud and stellar leakage, respectively; $b_{int\_1}$ and $b_{int\_2}$ are parameters that can be determined by a fit to detailed studies, $b_{int\_2}$ is dimensionless while $b_{int\_1}$ is in time$^{-1}$ unit.

Using Fig.6 from Defrère et al. 2010 at $\lambda$ = 10 µm, $D$ = 15 pc, $L$ = 1, $\Phi$ = 2 m, one gets the flux ratio $F_{SZ} / F_{st\_leaks}$ ~ 5, which allows the determination of $b_{int\_2}$. It reads: $b_{int\_2} \rho$ ~ 11.

---

[2] The IR stellar flux is $\propto kT_*$ (Rayleigh regime), bolometric is $\propto kT_*^4$, and the correction is $\propto kT_*^3$.

## 3.4 Required integration times

A signal to noise ratio, $S_i/N_i$, results from Eq.1, and Eq.2 or Eq.4. This $S_i/N_i$, must have a sufficient value to permit actual measurements, e.g., $S_i/N_i = 10$. Then, the needed integration time, $t_i$, results.

A possible observational sequence is as follows. In a target list, stellar systems are ordered by increasing time $t_i$ and studied one after the other until the total mission duration is reached. The number of accessible targets results.

- For *coronagraphs*, $t_i$ reads

$$t_i = c \; \Phi^{-2} \; \Delta\lambda^{-1} \; R_{pl}^{-4} \; L_i \; \rho(i) \; D_i^2 \qquad \text{(Eq.5)}$$

where $c$ is a parameter in time unit. It can be obtained by fitting the results of detailed models of coronagraphs, e.g., Lunine et al. 2009, or Levine et al. 2009.

- For *interferometers*, $t_i$ reads

$$t_i = d \left[ 1 + 11 \, L_i \; (T_* / T_\odot)^{-3} \; \Phi^2 \, D_i^{-2} \right]^{/2} \; \Phi^{-4} \; \Delta\lambda^{-1} \; R_{pl}^{-4} \; D_i^4 \qquad \text{(Eq.6)}$$

where $d$ is a parameter, in time unit, that can be determined in a similar way.

- For *both instruments*, the number of accessible targets during the mission lifetime e.g. 5 yrs, $N_{tot}$, is obtained by resolving:

$$\sum_{i=1}^{N_{tot}} t_i = 5 \, yrs \qquad \text{(Eq.7)}$$

Eq.5, 6, and 7, provide *explicit dependences* of the instrument capabilities upon quantities as $\Phi$ [m], $\Delta\lambda$ [µm], $R_{pl}$ [R$_{Earth}$], $L_i$ [L$_\odot$], $\rho_i$, $D_i$ [pc], and $\eta_{earth}$, and are exploited thereafter. They make the impact of these parameters easy to grasp.

## 3.5 Required signal to noise and spectral resolution

The upper parts of Fig.1 and 2 show the Earth spectrum as it would be seen by a distant observer using S/N = 10, $\lambda/\delta\lambda$ = 70 in the visible and 20 in the thermal IR. The bottom plot of these two figures shows the case of a twice larger S/N (x 4 integration time, no systematic noise) and a twice larger spectral resolution (x 2 integration time). The information content is significantly larger in these bottom plots.

This suggests that the higher spectral resolutions ($\lambda/\delta\lambda$ = 140 in the visible, and $\lambda/\delta\lambda$ = 40 in the mid-IR) should be included in future visible coronagraphs and mid-IR spectrometer.

We note that S/N = 10 for an Earth turns into S/N = 20 for a 1.4 $R_{earth}$ super-Earth, which would make the detection of molecular species significantly easier.

In the thermal IR, the spectrum shown in the top plot of Fig. 2 would allow $CO_2$ (> 10 $\sigma$) and possibly $O_3$ (~ 3 $\sigma$) to be detected. The bottom plot would allow the same species to be detected but without ambiguity for $O_3$ (~ 7 $\sigma$). In addition, one may argue for a marginal detection of the temperature inversion in the atmosphere from the shape of the 15 $\mu$m $CO_2$ band, and of the presence of the 7.8 $\mu$m $CH_4$ band (~ 2 $\sigma$). The footprint of $H_2O$ in the 6 - 8 $\mu$m domain could also enter the discussion. It appears as a broad deviation from a black body spectrum determined from the simulated data points, including noise, obtained in the equivalent of the Earth's atmospheric windows, 8.2 - 9.2 $\mu$m and 10.5 - 12.0 $\mu$m. This discussion suggests the need for the mid-IR spectrometer to reach a spectral resolution $\lambda/\delta\lambda$ = 40.

In the rest of the paper, the lower values for S/N (= 10) and $\lambda/\delta\lambda$ (= 70 or 20) are adopted. The corresponding limited quality of the spectra is not as severe as it may seem. If the spectra actually observed were similar to those of the top plots of Fig.1 and 2, the detection of $O_2$ in the visible (0.76 $\mu$m) and $CO_2$ in the thermal IR (15 $\mu$m) would be at least suspected. The observers' enthusiasm would be such that longer integration times (e.g., x 8) would be attributed to the corresponding planets, and better spectra obtained, *in the favorable case where systematic noise does not dominate the error budget,* a key point for attention when building the missions.

In conclusion, we think that the selected values are relevant to compare the capabilities of different instruments, but are probably not the final ones.

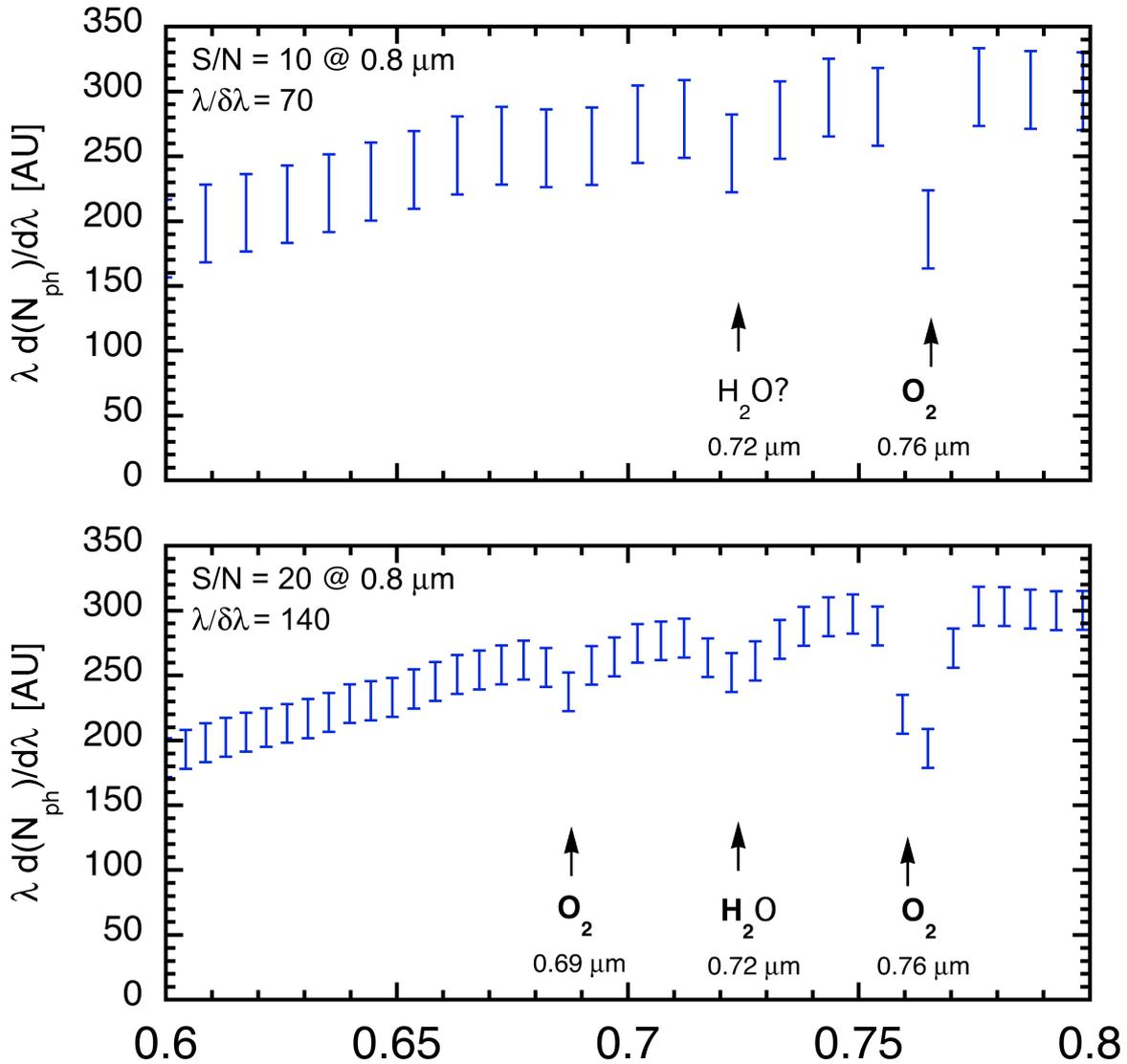

<u>Figure 1</u>: Earth spectrum in the 0.6 – 0.8 μm domain calculated from DesMarais et al. 2002 (Fig. 3 therein) in the case of a medium cloud coverage, for different S/N and spectral resolutions. The top plot corresponds to S/N = 10 (at 0.8 μm) and $\lambda/\delta\lambda$ = 70, and bottom plot to S/N = 20 (at 0.8 μm) and $\lambda/\delta\lambda$ = 140. The error bars indicate the intervals ± 1 σ. The top plot would allow only the $O_2$ A band to be detected (a single spectral point, at 4 σ relatively to the continuum). In the bottom plot, two species would be detected in three spectral bands: $O_2$ in its A and B bands, (8 σ and 3 σ, respectively); and $H_2O$ in its 0.72 μm band (3 σ). See text for implications.

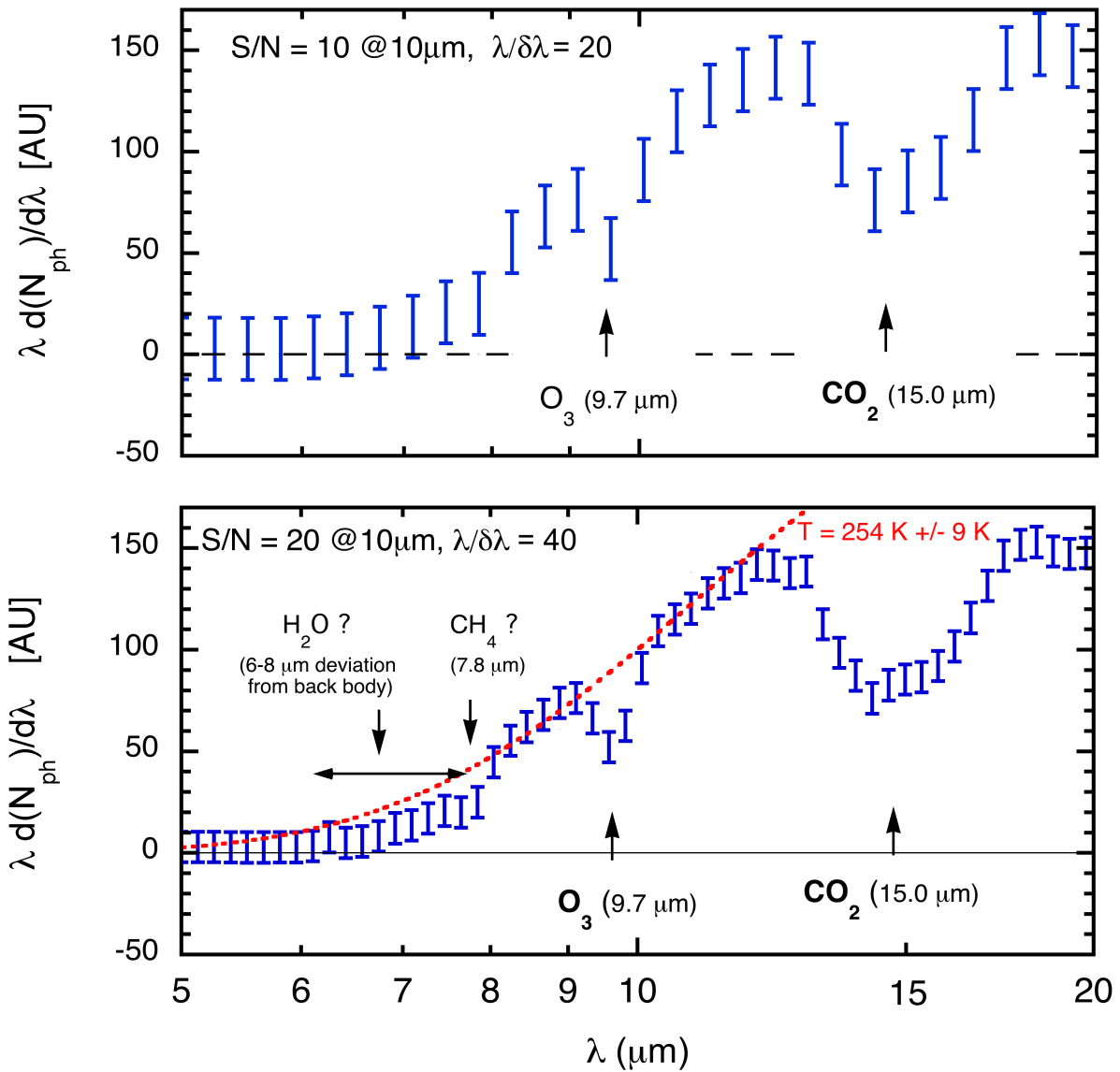

Figure 2: Same as Fig.1 but for the Earth emission in the thermal IR calculated from DesMarais et al. 2002 (Fig. 2 therein). Spectral resolution, λ/δλ, is constant over the 5 – 20 μm domain. Top plot corresponds to S/N = 10 (at 18 μm) and λ/δλ = 20, bottom plot to S/N = 20 (at 18 μm) and λ/δλ = 40. The top plot would allow $CO_2$ (> 10 σ) and possibly $O_3$ (~ 3 σ) to be detected. The bottom plot would allow the same species to be detected but without ambiguity for $O_3$ (~7 σ). In addition, one may argue for a marginal detection of the temperature inversion in the atmosphere from the shape of the 15 μm $CO_2$ band, and of the presence of the 7.8 μm $CH_4$ band (~ 2 σ). The footprint of $H_2O$ in the 6 - 8 μm domain could also enter the discussion. It appears as a broad deviation from a black body spectrum obtained in the equivalent of the Earth's atmospheric windows, 8.2 - 9.2 μm and 10.5 - 12.0 μm. This black body background (T = 254 K ± 9 K) is obtained by a two parameter fit (amplitude, and T) to the calculated spectral points plus an added gaussian noise realization with σ = N.

# 4. Coronagraphs

## 4.1 Double star problem

Can a stellar companion prevent the study of a target star, the companion being physically bound, or not (object in projection)?

In the vicinity of the target, the light from a companion is from its Airy rings and from its speckles due to defects in the mirrors. The former are deterministic and can be subtracted if the instrumental PSF is known, leaving only the quantum noise. The latter requires that the speckles can be reduced by a specific method, such as that proposed by Thomas et al. (2014).

The corresponding S/N ratio requirement leads to a constraint on the angular distance of the companion. The signal is the number of photo-electrons on the detector from a possible planet, $N_{pl}$, and a *minimum estimate of the noise* is the quantum noise from the Airy rings of the companion, $(N_{comp})^{1/2}$.

Using typical values for the nearest stars: $D$ = 3 pc, a companion with a R magnitude $R_{comp}$ = 3, a $\Phi$ = 2.4 m telescope, and an integration time of one day (1 d), a S/N of 10 requires an angular separation (Annex):

$$\theta_{comp} > 35.5 \ (D/3 \text{ pc})^{1.33} \ 10^{-0.13\,(R_{comp}-3)} \ (\Phi/2.4 \text{ m})^2 \ (t/1\text{d})^{-1/3} \qquad [\text{as}] \qquad \text{(Eq.8)}$$

Only stellar systems fulfilling this requirement are kept in the target list.

This is a *minimum requirement*, assuming full cancelation of the speckles due to the companion and the mirror defects by an appropriated system (Thomas 2014), probably an optimistic assumption.

The Hipparcos catalogue of nearby F, G, K, M stars is used in its 2007 version of data reduction (van Leeuwen, 2007), with stellar features (distance, luminosity...) as compiled by Jesus Maldonado (private communication) and Margaret Turnbull (2015, in preparation). Features of the 15 nearest FGK stars with a bright companion and resulting minimum integration time are given in Table 1. The key $\alpha$ Cen (A,B) multiple system would require a special effort, 3.6 $10^{-2}$ yr = 13 days of continuous integration, whereas the other ones, with the exception of 70 Oph A, should be observable (see also Sect. 6.1).

**Table 1**

Nearby FGK double stars[a]

| target star | $\alpha$ Cen B | $\varepsilon$ Eri | 61 Cyg B | Procyon A | 70 Oph A |
|---|---|---|---|---|---|
| $D$ [pc] | 1.29 | 3.21 | 3.5 | 3.5 | 5.1 |
| $R_{comp}$ [mag] | 0.0 | 17.3 | 5.2 | 10.8 | 6.2 |
| $\theta_{comp}$ [as] | 10 (in 2030) | 17.1 | > 20 as | 4.7 | 6.2 |
| $t_{integ}$[a] [yr] | 3.6e-2 | << 1.0e-3 | < 4.1e-3 | 2.2e-3 | 2.3e-1 |

[a] for a 2.4 m mirror

## 4.2 Orbital and orientation problems

For face-on circular orbits, the signal from a planet is continuously that at full elongation, that from an illuminated half disc.

For inclined orbits the situation is less favorable. Planets close to the IWA at full elongation spend only a small fraction of their time outside of it along their orbit. For $\alpha$ = 60°, the mean value of inclination angle for randomly oriented orbits, this fraction is only 31% for a planet whose full elongation is 1.1 times the IWA. Without a prior detection determining the planetary mass, orbit and ephemerides, four visits would be needed. For planets closer to their stars, e.g., 1.05 x IWA, the fraction would be 23%, requiring five visits. A compromise could be to limit the search to three or four visits per target, accepting not being exhaustive. However, the planetary mass would remain unknown, with the corresponding lack in our understanding of the physics at the planetary surface (see Sect.9).

In most of the next sections, a prior detection of planets in the HZ of the nearby stars is assumed.

## 4.3 Light rejection by a coronagraph

Following Lunine et al. (2009) (= Lunine et al. 2009) (Fig.11.2 therein), a simplified dependence of the light transmission as a function of the angle between source and instrument axis is used (Fig.3), with transmission $\rho$:

$$\rho(\theta_i) = 10^{-10\,\theta_i/\overline{IWA}} \quad \text{for } \theta_i/\overline{IWA} < 1 \qquad\qquad (Eq.9)$$
$$= 10^{-10} \qquad\quad \text{for } \theta_i/\overline{IWA} > 1$$

where $\theta_i$ is the angular distance of the HZ of star ($i$), the star-planet angular distance when planet is at full elongation, $\theta_i = \theta_{HZ,i}$, $= a_{HZ,i}/D_i$, and $\overline{IWA}$ is the Inner Working Angle of the coronagraph, in radian.

For a system at distance $D_i$, the transmission reads

$$\rho(D_i) = 10^{-10\,/(D_i/D_{ci})} \quad \text{for } D_i/D_{ci} < 1 \qquad\qquad (Eq.10)$$
$$= 10^{-10} \qquad\quad \text{for } D_i/D_{ci} > 1$$

where $D_{c,i}$ is the critical distance of system (i), $D_{c,i} = a_{HZ,i}\cdot\overline{IWA}^{-1}$. At $\lambda = 0.8\,\mu m$, it reads: ($D_{c,i}/1pc$) = 6.06 ($\Phi/1m$) IWA$^{-1}$ $a_{HZ,i}$ $L_i^{1/2}$ [pc], with IWA in $\lambda/\Phi$ unit, and $a_{HZ,i}$ in AU $L_i^{1/2}$ unit.

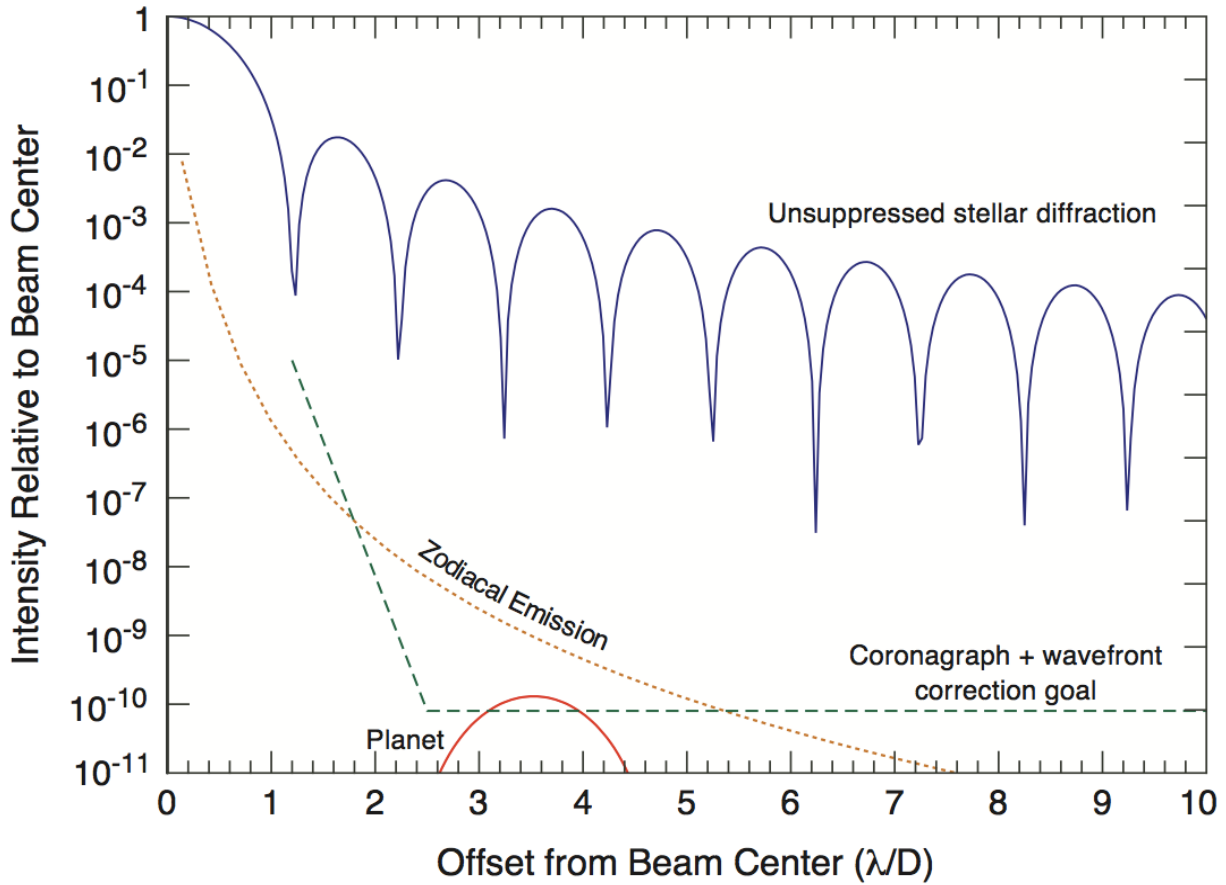

Figure 3: schematic light transmission by an optical system vs the angular distance of the source from its optical axis for a telescope free from central obscuration (full line = Airy function), and for a telescope with a coronagraph having a transmission of $10^{-10}$ at 2.5 $\lambda/\Phi$ (dashed line). Solar zodiacal light (dotted line) is stable and can be subtracted, but produces quantum noise. Exo-zodiacal light is not represented. The PSF of a possible Earth-like planet is represented by a full line at the low part of the figure (from Lunine et al. 2009).

## 4.4 Calibrating the integration time, value of *c*

A detailed study by Lunine et al. 2009 (Table 2 therein) states that a coronagraph with a 5 m telescope, IWA = 3 $\lambda/\Phi$, or 100 mas at $\lambda$ = 0.8 μm, searching for Earth-size planets ($R_{pl}$ = 1.0) at $a_{HZ}$ = 1.0 $L_i^{1/2}$ [AU] with a spectral resolution suitable for detection ($\lambda/\Delta\lambda \sim 2$), can investigate 44 stars, in 5 yrs.

The parameter *c* can be determined by using Eq.5 and Eq.7, and that requirement, $N_{tot}$ = 44 for a 5 yr mission. Table 2 gives the integration times divided by *c*, $t_i/c$, and their sum, S($t_i/c$), for stars in the target list, until rank 44 is reached (Hip15371, with S($t_i/c$) = 1.78 × 10⁻⁸). The value of *c* results:

$$c = 5 \text{ yrs} / 1.78 \times 10^{-8} = 2.81 \times 10^{8} \text{ yrs}$$

**Table 2**

Integration times divided by *c* for FGK stars[a]

| HIP | Common | D[pc] | V | $L_{bol}/L_0$ | SpType | ρ | ti /c | star # | S(ti/c) |
|---|---|---|---|---|---|---|---|---|---|
| 71681 | α Cen B | 1.29 | 1.35 | 0.525 | K0V | 1.00e-10 | 6.99e-12 | 1 | 6.99e-12 |
| 104217 | 61 Cygni B | 3.50 | 5.95 | 0.097 | K7.0V | 9.97e-10 | 9.49e-12 | 2 | 1.65e-11 |
| 104214 | 61 Cygni A | 3.50 | 5.20 | 0.134 | K5.0V | 1.00e-10 | 1.31e-11 | 3 | 2.96-11 |
| 71683 | α Cen A | 1.29 | -0.01 | 1.611 | G2.0V | 1.00e-10 | 2.14e-11 | 4 | 5.10-11 |
| (…) | | | | | | | | | |
| 10644 | NA | 10.78 | 4.86 | 1.213 | G0V | 1.00e-10 | 1.13e-09 | 43 | 1.67e-08 |
| 15371 | NA | 12.03 | 5.24 | 1.034 | G2V | 2.91e-09 | 1.20e-09 | **44** | **1.78e-08** |
| 80337 | NA | 12.78 | 5.37 | 1.034 | G3/5V | 9.16e-09 | 1.35e-09 | 45 | 1.91e-08 |
| (…) | | | | | | | | | |

[a] Conditions are those by Lunine et al. 2009. An exhaustive list, up to rank 200, is available in electronic form

*The uncertainty on the model predictions* can be estimated by using the parameter value determined above, *c* = 2.81 × 10⁸ yrs, and applying the model to coronagraphs with different features, as those studied by Levine et al. 2009, and comparing the predictions by the model to those by the detailed studies. For the flagship TPF-C instrument (2002), a 8 m mirror working at IWA = 4 $\lambda/\Phi$, and $R_{pl}$ = 1.0 planet in the HZ, the model predicts that 71.6 stellar systems can be investigated in 3 yrs, whereas Levine et al. 2009 find 73 systems. For Φ = 2.5 m, IWA = 2.5λ/Φ, 3 yr mission, the model finds 14.3 systems, instead of 16 (Levine et al. 2009). *The similarity between these numbers indicates the robustness of the model.*

## 4.5 Distinguishing planet and exozodiacal disc

It is well known that the presence of exozodiacal dust around a star can hamper the direct detection and characterization of telluric planets in the habitable zone (e.g., Beichman et al. 2006, Roberge et al. 2012). Its light produces a signal much stronger than the planetary signal, 350 times in the case of 1 zodi (the zodiacal light luminosity of the solar system) at 10 μm (Defrère et al. 2010), and a similar value in the visible. The planetary signal must be disentangled from it by studying the image of the system, at low spatial resolution (a few resolution elements, λ/Φ, in the image of the planet-disc system). The impact is twofold: (1) as a source of noise (Brown et al. 2005), and (2) as a source of confusion (e.g., Defrère et al. 2012).

Roberge et al. (2012) published the first report by NASA's ExoPAG working group on that problem, and concluded that, for a 2 m-class telescope, the required observing time increases dramatically with the level of exozodiacal light, *even from zero to one zodi* (Fig. 2, therein). They also point out that this has all the more consequences on the capability of a 2 m coronagraph than $\eta_{earth}$ is small. Obviously, if all stars have very bright exozodiacal disks, the value of $\eta_{earth}$ does not really matter since it will be extremely difficult to detect any planet anyway.

Very little is currently known about the prevalence of exozodiacal dust around nearby main sequence stars. Based on Keck Nuller observations of a sample of 20 solar-type stars with no far IR excess previously detected (i.e., no detectable outer dust reservoir), a recent study by Mennesson et al. (2014) concludes with high confidence (95%) that the median level of exozodiacal dust around such stars is below 60 times the solar value. This state-of-the-art sensitivity to exozodiacal dust is however much too high to decide what is the fraction of stars with sufficiently low dust levels to enable the detection of telluric planets in the HZ. This requires a better knowledge of the faint-end of the exozodiacal dust luminosity function, which should be obtained soon by the LBTI instrument (Defrère et al. 2015). Therefore, given the lack of knowledge on this topic, we make *the arbitrary assumption* in this study that the level of exozodiacal emission is sufficiently low, and ignore it, leaving the study of the impact of this spurious light on coronagraphs for a more detailed paper.

## 4.6 Performances of an affordable coronagraph

What could be an affordable coronagraph? Within the present and mid-future funding possibilities, a 2.4 m telescope seems more likely than a 5 m one. Recently, NASA acquired two such telescopes and is considering the implementation of a coronagraph on one of them, WFIRST-AFTA (WFIRST 2014). Unfortunately, it has a central obstruction and six struts for holding the secondary mirror, making it less suitable than an off-axis telescope. This translates into a contrast that cannot reach the required $10^{-10}$ value, but makes it still valuable for characterizing larger planets, an opportunity not to miss.

An optimized off-axis telescope, with the same mirror size, seems affordable in the mid-future, at mass and cost similar to that of WFIRST-AFTA. The capabilities of a coronagraph on a 2.4 m telescope are used, with a $10^{-10}$ transmission, IWA = 2.5 $\lambda/\Phi$ or 170 mas at $\lambda = 0.8$ µm, assuming that progress in the future will make this performance reachable. A worksheet similar to Table 2 is made with the above values and those from Sect.2 and 4.3: $a_{HZ,I} = 1.30\ L_i^{1/2}$ [AU], $D_{c,i} = 7.56\ L_i^{1/2}$ [pc], $R_{pl} = 1.5$, $\lambda/\Delta\lambda = 70$ suitable for spectroscopy searching for the 0.70 µm and 0.77 µm $O_2$ bands (DesMarais et al. 2002), and $c = 2.81 \times 10^8$ yrs (Table 3).

The impact of $\eta_{earth}$ is estimated as follows. A target list is used with stars ranked by increasing integration time $t_i$. The total time up to rank ($i$), S($t_i$), is calculated. If $\eta_{earth}$ is less than 100% only one star out of $1/\eta_{earth}$ has a planet. *If the stars with relevant planets are known in advance*, it could be possible to go deeper in the list, leaving aside the non-relevant stars. For a given mission duration it will be equivalent to investigate all stars, up to number $N_{virt}$, in a virtual mission that would last longer, 5 yrs yrs / $\eta_{earth}$, but out of which only $N_{virt}$ x $\eta_{earth}$ have planets.

If all stars had the same observation times, the value of $\eta_{earth}$ would have no impact on the number of studied planets: $N_{pl} = N_{pl}\ (1/\eta_{earth})$ x $\eta_{earth}$ In reality, the integration time increases steeply with the rank in the lists (Table 2 and 3), and lower values of $\eta_{earth}$ lead to significantly lower numbers of planets. This reasoning would be exact for objects in large numbers. For objects in small number it has the limitations of small number statistics.

*Previous identification of the suitable stars* is assumed, and the ephemerides determined, so that the whole mission time can be spent on spectroscopy. This is of special importance if $\eta_{earth}$ is significantly smaller than 100% so that many nearby stars have no suitable planet. The spectroscopic mission would not lose time on them (Sect.9).

Using Table 3, one finds that for $\eta_{earth} = 100\%$, a 5 yr mission, and $R_{pl} = 1.5$, 8 planets can be studied in spectroscopy around the 8 first stars; for $\eta_{earth} = 10\%$ the virtual mission lasts 50 yrs, $N_{virt}$ is 15, and 1.5 planet can be studied; for $\eta_{earth} = 1\%$, $N_{virt}$ is 34, and $N_{pl} = 0.34$.

**Table 3**
Possible target list for a 2.4 m coronagraph

| Hipparcos | Name | D (pc) | V | L_bol (L_Sun) | Sp Type | rho | ti (yr) | (i) | S_(ti) (yr) |
|---|---|---|---|---|---|---|---|---|---|
| 71681 | alf Cen B | 1.29 | 1.35 | 0.525 | K0V | 1.00E-10 | 5.87E-02 | 1 | 5.87E-02 |
| 71683 | alf Cen A | 1.29 | -0.01 | 1.611 | G2.0V | 1.00E-10 | 1.8E-01 | 2 | 2.39E-01 |
| 108870 | eps Indi A | 3.62 | 4.69 | 0.227 | K4V | 1.16E-10 | 2.32E-01 | 3 | 4.71E-01 |
| 16537 | eps Eri | 3.21 | 3.71 | 0.351 | K2.0V | 1.00E-10 | 2.44E-01 | 4 | 7.15E-01 |
| 8102 | tau Ceti | 3.65 | 3.49 | 0.519 | G8.5V | 1.00E-10 | 4.65E-01 | 5 | 1.18E+00 |
| 19849 | omi 2 Eri | 4.98 | 4.43 | 0.425 | K0.5V | 1.30E-10 | 9.21E-01 | 6 | 2.10E+00 |
| 88601 | 70 Oph | 5.10 | 4.03 | 0.681 | K0V | 1.00E-10 | 1.19E+00 | 7 | 3.29E+00 |
| 15510 | 82 Eri | 6.04 | 4.26 | 0.691 | G8.0V | 1.00E-10 | 1.70E+00 | 8 | 4.99E+00 |
| 3821 | eta Cas A | 5.94 | 3.45 | 1.312 | G3V | 1.00E-10 | 3.12E+00 | 9 | 8.11E+00 |
| 99240 | del Pav | 6.11 | 3.53 | 1.332 | G8.0IV | 1.00E-10 | 3.34E+00 | 10 | 1.15E+01 |
| 37279 | Procyon A | 3.51 | 0.40 | 7.118 | F5IV-V | 1.00E-10 | 5.88E+00 | 11 | 1.73E+01 |
| 61317 | bet CVn | 8.44 | 4.24 | 1.268 | G0V | 1.00E-10 | 6.07E+00 | 12 | 2.34E+01 |
| 1599 | ___ | 8.59 | 4.23 | 1.333 | G0V | 1.00E-10 | 6.61E+00 | 13 | 3.00E+01 |
| 64394 | ___ | 9.13 | 4.24 | 1.477 | G0V | 1.00E-10 | 8.27E+00 | 14 | 3.83E+01 |
| 105858 | ___ | 9.26 | 4.22 | 1.545 | F7V | 1.00E-10 | 8.91E+00 | 15 | 4.72E+01 |
| 89937 | chi Dra | 8.06 | 3.56 | 2.141 | F7Vvar | 1.00E-10 | 9.34E+00 | 16 | 5.65E+01 |
| 86974 | ___ | 8.31 | 3.41 | 2.776 | G5IV | 1.00E-10 | 1.29E+01 | 17 | 6.94E+01 |
| 22449 | 1 Ori | 8.07 | 3.17 | 2.997 | F6V | 1.00E-10 | 1.31E+01 | 18 | 8.25E+01 |
| 104214 | 61 Cyg A | 3.50 | 5.20 | 0.134 | K5.0V | 1.22E-08 | 1.34E+01 | 19 | 9.59E+01 |
| 27072 | ___ | 8.93 | 3.59 | 2.514 | F7V | 1.00E-10 | 1.35E+01 | 20 | 1.09E+02 |

| 2021 | bet Hyi | 7.46 | 2.82 | 3.702 | G1IV | 1.00E-10 | 1.38E+01 | 21 | 1.23E+02 |
|---|---|---|---|---|---|---|---|---|---|
| 32349 | Sirius A | 2.63 | -1.44 | 30.483 | A1.0V | 1.00E-10 | 1.42E+01 | 22 | 1.37E+02 |
| 14632 | ___ | 10.54 | 4.05 | 2.355 | G0V | 1.00E-10 | 1.76E+01 | 23 | 1.55E+02 |
| 17378 | ___ | 9.04 | 3.52 | 3.374 | K0IV | 1.00E-10 | 1.85E+01 | 24 | 1.74E+02 |
| 97649 | Altair | 5.12 | 0.76 | 10.649 | A7IV-V | 1.00E-10 | 1.87E+01 | 25 | 1.92E+02 |
| 12777 | ___ | 11.13 | 4.10 | 2.418 | F7V | 1.00E-10 | 2.01E+01 | 26 | 2.12E+02 |
| 96100 | sig Dra | 5.75 | 4.67 | 0.436 | G9.0V | 2.10E-09 | 2.04E+01 | 27 | 2.33E+02 |
| 27913 | chi Ori | 8.66 | 4.39 | 1.166 | G0V | 3.78E-10 | 2.22E+01 | 28 | 2.55E+02 |
| 78072 | ___ | 11.25 | 3.85 | 3.134 | F6V | 1.00E-10 | 2.67E+01 | 29 | 2.82E+02 |
| 57757 | ___ | 10.93 | 3.59 | 3.792 | F9V | 1.00E-10 | 3.04E+01 | 30 | 3.12E+02 |
| 109176 | ___ | 11.73 | 3.77 | 3.631 | F5V | 1.00E-10 | 3.36E+01 | 31 | 3.46E+02 |
| 7513 | ___ | 13.49 | 4.09 | 3.636 | F8V | 1.00E-10 | 4.45E+01 | 32 | 3.90E+02 |
| 116771 | ___ | 13.71 | 4.13 | 3.698 | F7V | 1.00E-10 | 4.67E+01 | 33 | 4.37E+02 |
| 61941 | ___ | 11.68 | 3.44 | 5.276 | F1V | 1.00E-10 | 4.84E+01 | 34 | 4.85E+02 |
| 102485 | ___ | 14.68 | 4.13 | 4.056 | F5V | 1.00E-10 | 5.87E+01 | 35 | 5.44E+02 |
| 81693 | ___ | 10.72 | 2.81 | 7.732 | F9IV | 1.00E-10 | 5.97E+01 | 36 | 6.04E+02 |
| 70497 | ___ | 14.53 | 4.04 | 4.419 | F7V | 1.00E-10 | 6.27E+01 | 37 | 6.66E+02 |
| 16852 | ___ | 13.96 | 4.29 | 3.292 | F8V | 1.50E-10 | 6.46E+01 | 38 | 7.31E+02 |
| 113368 | Fomalhaut | 7.70 | 1.23 | 16.466 | A3V | 1.00E-10 | 6.57E+01 | 39 | 7.97E+02 |
| 59199 | ___ | 14.94 | 4.02 | 4.558 | F0IV/V | 1.00E-10 | 6.84E+01 | 40 | 8.65E+02 |

Notes:  A list up rank 100 is available in electronic form
[a] required integration time for target (i) and $R_{pl}$ = 1.5
[b] cumulative time for targets 1,2, …, i

It is noticeable that among these best 100 systems, *there is no M star*. The first M star is Proxima Cen (D = 1.30 pc) with the rank 210; Barnard's star (D = 1.83 pc) is 224[th]. This is due to the low efficiency of coronagraphs for planets inside the IWA, which is the case for the HZ of low luminosity stars. Coronagraphic transmission in the HZ is $1.9 \times 10^{-2}$ and $3.1 \times 10^{-3}$ for the two quoted M stars, far from the $10^{-10}$ value outside the IWA.

This procedure can be repeated for different values of $\eta_{earth}$ and planetary radius. The number of planets that can be studied in spectroscopy results as a function of $\eta_{earth}$ (Table 4 and Fig.4). When the size distribution of planets in the HZ and $\eta_{earth}$ will be known, a precise estimate of the number of accessible planets will be possible.

The dependence of $N_{pl}$ on $\eta_{earth}$ can be fitted by a power law with an exponent $\beta_{coron} = 0.71$.

Table 4 and Fig.4 are one of the *major results of the present paper*. They give the number of planets that can be studied in spectroscopy, with a built-in coronagraph. For $\eta_{earth}$ = 10%, it is between 1 and 2.3 according the planetary radius, $N_{pl} = 1.5$ for $R_{pl} = 1.50$. This is very small number statistics and it is therefore affected by the corresponding uncertainties.

**Table 4**

Number of planets accessible to a
2.4 m coronagraph

| $\eta_{earth}$ (%) | $R_{pl} = 1.0$ | $R_{pl} = 1.5$ | $R_{pl} = 2.0$ |
|---|---|---|---|
| 1 | 0.19 | 0.34 | 0.48 |
| 3 | 0.39 | 0.75 | 1.05 |
| 10 | 1.00 | 1.50 | 2.30 |
| 30 | 2.10 | 3.30 | 4.80 |
| 100 | 5.00 | 8.00 | 11.00 |

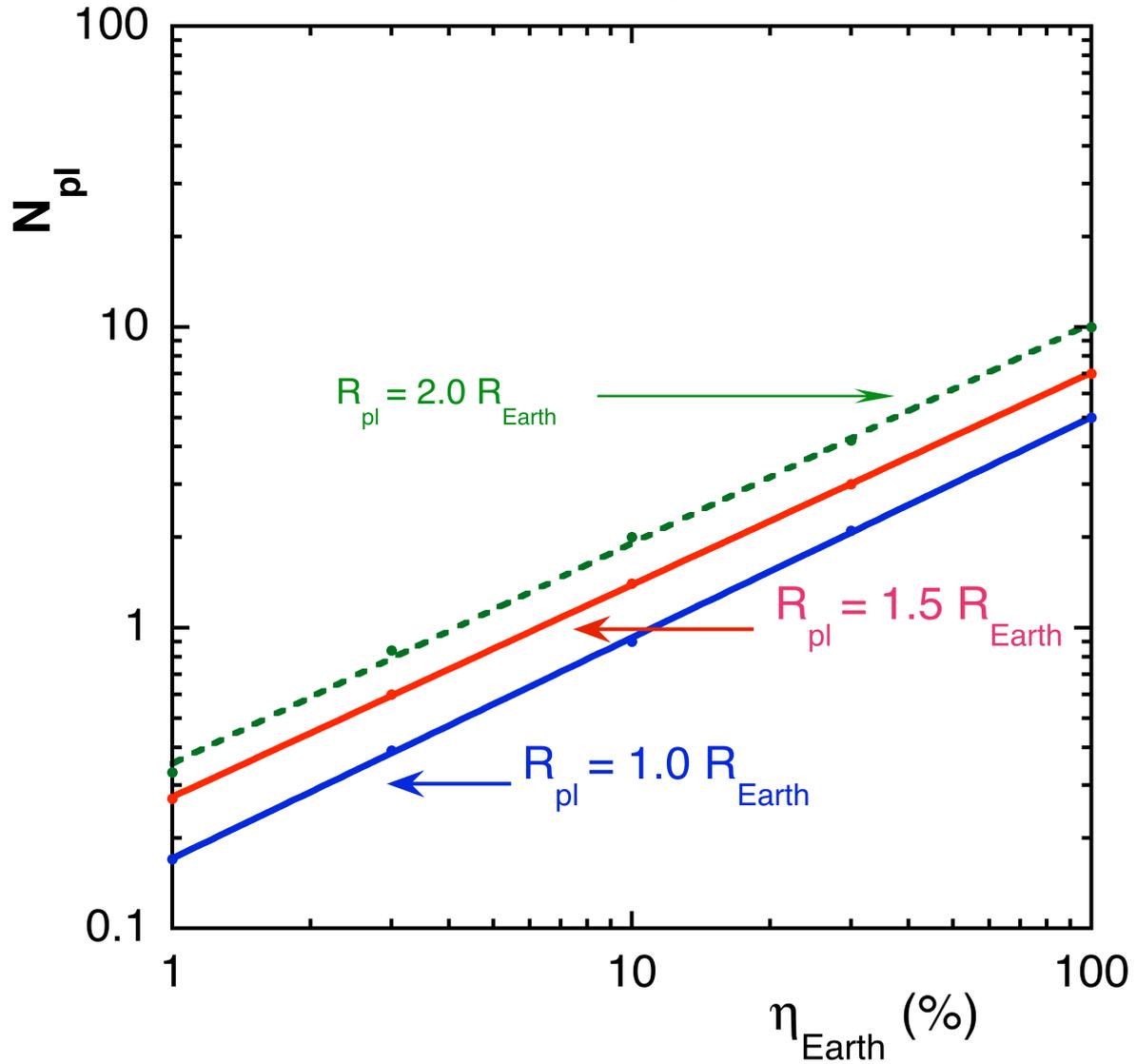

**Figure 4:** Number of planets located in the habitable zone that can be studied in spectroscopy ($\lambda/\Delta\lambda \sim 70$) with a IWA = 2.5 $\lambda/\Phi$ coronagraph on a 2.4 m telescope, versus $\eta_{earth}$, for different values of the planetary radius, under the major hypothesis that exozodiacal light does not prevent planet detection. The case of $R_{pl} = 2.0$ $R_{Earth}$ planets is represented with a dashed line to indicate that they are probably not rocky (Rogers 2015, Fig.2 therein). Data points are calculated on target lists and are rather well aligned on power law lines with a $\beta_{coron} = 0.71$ exponent (lines).

## 4.7 Possible detection by the coronagraph

Although there would be major disadvantages to a situation where suitable planets are not previously identified (Sect. 9), one can estimate how many stars could be investigated by the coronagraph, searching for suitable planets, and later performing spectroscopy on the detected objects. Assuming that this could be done with 3 visits to limit the phase problems (Sect. 4.2), and a spectral resolution $\lambda/\Delta\lambda = 4$, the proportionality of the required time per visit to $\Delta\lambda$ (Eq. 5) gives the observing time. Explicitly, if 2.5 yrs are dedicated to the detection, the equivalent time in Table 3 is: 2.5 yrs x (1/3) x 70/4 = 14.6 yrs. The *coronagraph could investigate targets up rank 11, Procyon A* (3.51 pc).

For $\eta_{solar} = 10\%,$ the 2.5 yrs left would correspond to a virtual mission of 25 yrs, which exceeds cumulated times for rank 11. These 2.5 yrs are then sufficient to *study in spectroscopy 11 x 0.10 = 1.1 planet*, with the high risks of statistics on very small numbers, and the hypothesis on the level of exozodiacal light.

# 5. Starshade

A starshade is an interesting option for direct imaging, because it can suppress stellar light to a very high level, e.g., a few $10^{-11}$ transmission outside its IWA, the remaining light being not due to the defects of the collecting telescope but to those of the starshade, which is submitted to variable sunshine.

Its IWA is driven by wavelength, required rejection, shade size and associated telescope-shade distance, but not by the telescope features. The size of the latter determines the required integration time and the capability to separate the planetary image from the exozodiacal cloud. The Interim Report on Exo-S by NASA (Seager et al. 2014) provides an estimate of the capabilities of an affordable starshade.

In that report, the capabilities are calculated for the detection of earth-like planets located at 1.0 $L^{1/2}$ AU from their stars, and $R_{pl} = 1.0$, $\lambda = 0.6$ µm, $\lambda/\Delta\lambda = 7$, $S/N = 3$. The integration times are estimated for the best 20 stars ranked in a target list (Table 3.2-1 and 3.3-1, therein). The total integration time of that program is 68 days, and detections have a mean completeness of about 40%.

The calculation can be transposed to the spectroscopy of the identified planets. For spectroscopy in the 0.6 – 0.8 µm band ($O_2$ A-band at 0.76 µm) with $\lambda/\Delta\lambda = 70$, S/N = 10, with a 2.4 m telescope, on $R_{pl} = 1.0$, 1.5, or 2.0 planets in the HZ, (1.30 renormalized AU). Assuming quantum noise only, a total integration time of $\leq 350$ days is found for $\eta_{earth}$ = 100%, a situation where all the 20 targets could be studied. The 5 yr duration of the mission is sufficient for data transmission and star retargeting, as described in the Interim report. The limitation of such a mission comes from the IWA permitted by the 34 m starshade, not by the size of the telescope even for spectroscopy, and the extraction of the planetary image from that of the exozodiacal disc, which depends on the telescope size. With a 2.4 m telescope these limitations are similar to those for a built-in coronagraph (Sect. 4.5).

Table 5 shows the target list for the instrument proposed by Seager et al. (2014) and a 2.4 m mirror. Systems are ranked by decreasing angular distances for planets located at 1.3 $L^{1/2}$ AU at full elongation, until the IWA of the instrument (115 mas) is reached, which determines the cut-off of the list. Whatever the value of $\eta_{earth}$, a maximum of 20 stellar systems can be investigated with a 34 m starshade, at least in the approximation where the integration time increase prohibitively when the IWA is reached.

**Table 5**

Possible target list for a starshade

| Hip | Common | D (pc) | V | L_bol | Sp. type | Sep 1.3AU (mas) | Star # |
|------|---------|--------|-------|-------|----------|---------|--------|
| 17378 | GL 150 | 9.04 | 3.52 | 3.37 | K0IV | 272 | 1 |
| 86974 | GL 695 | 8.31 | 3.41 | 2.78 | G5IV | 269 | 2 |
| 8102 | tau Ceti | 3.65 | 3.49 | 0.52 | G8.5V | 264 | 3 |
| 3821 | eta Cas A | 5.94 | 3.45 | 1.31 | G3V | 258 | 4 |
| 99240 | del Pav | 6.11 | 3.53 | 1.33 | G8.0IV | 253 | 5 |
| 27072 | GL 216A | 8.93 | 3.59 | 2.51 | F7V | 238 | 6 |
| 14632 | GL 124 | 10.54 | 4.05 | 2.36 | G0V | 195 | 7 |
| 12777 | GL 107A | 11.13 | 4.10 | 2.42 | F7V | 187 | 8 |
| 15510 | 82 Eri | 6.04 | 4.26 | 0.69 | G8.0V | 184 | 9 |
| 1599 | GL 17 | 8.59 | 4.23 | 1.33 | G0V | 180 | 10 |
| 105858 | GL827 | 9.26 | 4.22 | 1.54 | F7V | 180 | 11 |
| 61317 | beta CVn | 8.44 | 4.24 | 1.27 | G0V | 179 | 12 |
| 64394 | GL 502 | 9.13 | 4.24 | 1.48 | G0V | 178 | 13 |
| 19849 | omi 2 Eri | 4.98 | 4.43 | 0.42 | K0.5V | 175 | 14 |
| 77257 | GL 598 | 12.12 | 4.41 | 2.22 | G0Vvar | 165 | 15 |
| 96100 | sig Dra | 5.75 | 4.67 | 0.44 | G9.0V | 154 | 16 |
| 64924 | 61 Vir | 8.56 | 4.74 | 0.87 | G5V | 147 | 17 |
| 23693 | GL 189 | 11.65 | 4.71 | 1.54 | F6/7V | 143 | 18 |
| 57443 | GL 442A | 9.22 | 4.89 | 0.86 | G3/5V | 135 | 19 |
| 29271 | GL 231 | 10.20 | 5.08 | 0.91 | G6V | 125 | 20 |

The number of accessible planets results (Table 6, and Fig.5). *For $\eta_{earth}$ = 10%, $R_{pl}$ = 1.50, this number is slightly larger than for the in-built coronagraph, 2.0 instead of 1.5.*

*It would decrease to 0.8 planet if an independent detection is not performed beforehand*, due to the low completeness of the detection by the starshade (about 40%, Seager et al. 2014).

Turnbull et al. (2012) presented a preliminary discussion of starshades within a broader range of parameters (mirror diameter, shade diameter...), including a 4 m mirror and a 50 m shade, that could detect at least one Earth-like planet, possibly three, in the HZ of its star, if $\eta_{earth}$ is larger than 10%.

**Table 6**

Number of planets accessible for a 34 m star shade
and a 2.4 m telescope

| $\eta_{earth}$ (%) | $R_{pl}$ = 1.0 | $R_{pl}$ = 1.5 | $R_{pl}$ = 2.0 |
|---|---|---|---|
| 1 | 0.21 | 0.21 | 0.21 |
| 3 | 0.63 | 0.63 | 0.63 |
| 10 | 2.10 | 2.10 | 2.10 |
| 30 | 6.30 | 6.30 | 6.30 |
| 100 | 21.00 | 21.00 | 21.00 |

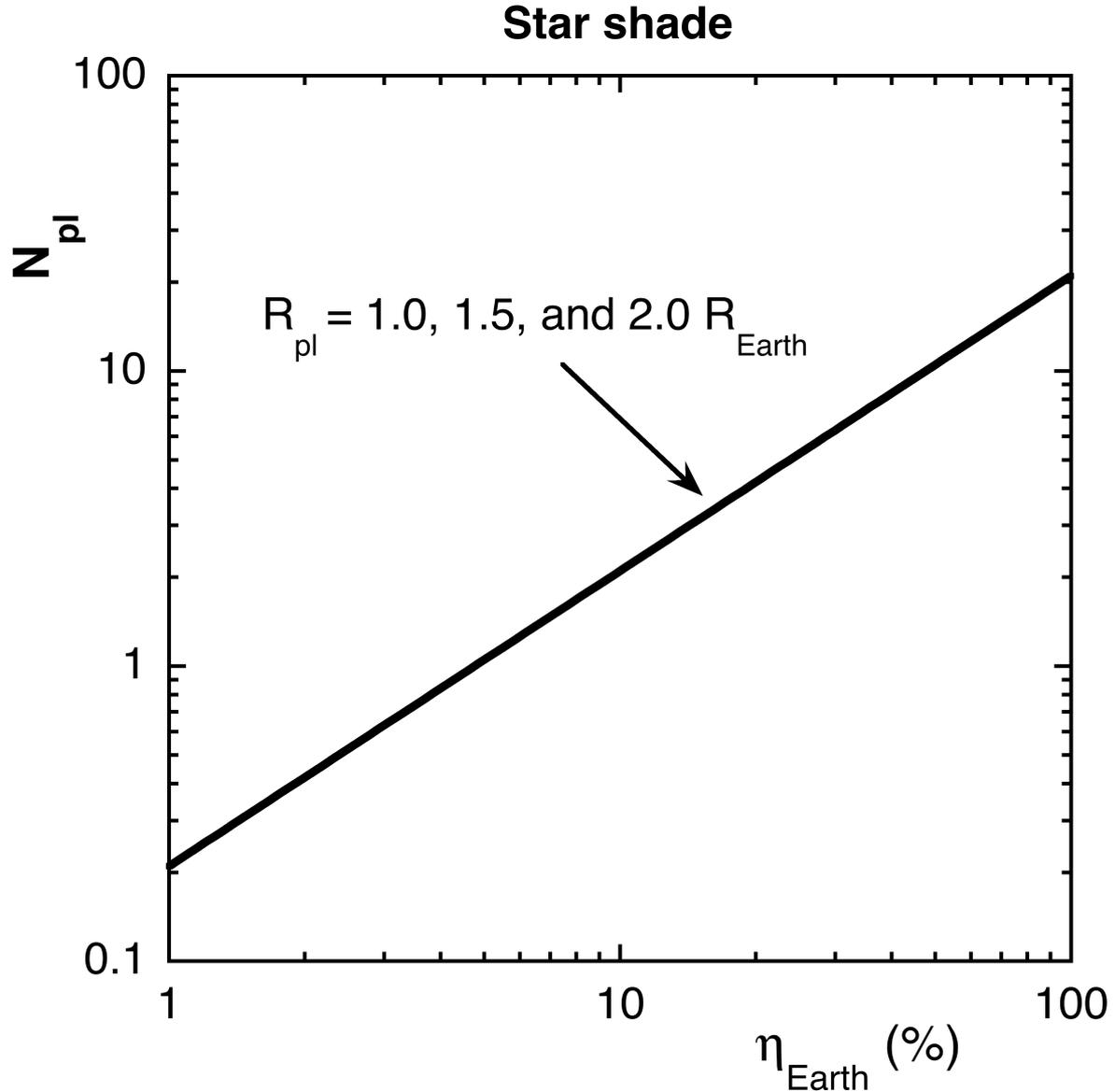

**Figure 5:** Number of planets located in the HZ that can be studied in spectroscopy ($\lambda/\Delta\lambda = 70$), with a 34 m starshade and a 2.4 m telescope, as a function of $\eta_{earth}$, assuming that the exozodiacal light of the target stars are low enough not to prevent the detection of telluric exoplanets in the HZ (see Sect. 4.5). The limitation of that number is driven by the inner working angle (IWA) of the shade, and independent of the planetary radius. However, the larger the planet is, the larger the S/N ratio. The planet is assumed at full elongation and located at 1.30 $L^{1/2}$ AU from its star. Data points follow a power law with a $\beta_{shade} = 1.0$ exponent. A beforehand detection of stars with suitable planet(s) is assumed, e.g., by astrometry. If detection is done by the starshade itself, all numbers must be divided by 2.5, due to the low completeness of the detection by the starshade (about 40%, Seager et al. 2014).

# 6. Nulling interferometers

## 6.1 Double star problem

Nulling interferometers inject the stellar light from their collecting mirrors into a single mode fiber (SMF) centered on the target star. The impact of a stellar companion depends on how much of its light is also injected in the fiber. A possible requirement is that this light is less than the target leaks, $10^{-5}$ times the target flux. Guyon (2002) gives the efficiency of the coupling of a source, $E_{coupl}$, as a function of its angular distance to the SMF axis. For an interferometer with $\Phi = 0.75$ m mirrors, at the most demanding wavelength $\lambda = 18$ μm, and with $\Delta m_{18}$ the difference between the companion and target magnitudes at 18 μm, the condition on the ratio, $r$, of light from the companion to that from the target into the SMF is: $r = 10^{-0.4 \, \Delta m\_18} \times E_{coupl}(\theta_{comp}) < 10^{-5}$. The upper envelop of the $E_{coupl}(\theta_{comp})$ curve by Guyon (2002, Fig.1) can be fit by a second degree polynomial, and the condition on the angular distance of the companion reads:

$$\theta_{comp} > 20 - 0.96 \times \Delta m_{18} - 0.055 \times \Delta m_{18}^2 \qquad [as] \qquad (Eq.11)$$

Only binary stars with companion fulfilling that requirement are kept in the target list, with the exception of the $\alpha$ Cen system for which an adequate integration time should allow to push down that limit.

It should be noted that the case of our nearest neighbors, the triple system Alpha Cen A, B, and Proxima Cen, is special. These stars are at 1.3 pc, significantly nearer than what a uniform density of stars would give (if the stellar density was uniform, the first solar-type star[3], would be at 2.6 pc). As a consequence, this triple system is at the very top of all target lists: astrometry, coronagraphs, and nulling interferometers. Two questions remain: (i) Will all these instruments be able to separate the A and B components? (ii) Do these stars have habitable planet(s)?

## 6.2 Orbital and orientation problems

For interferometers in the thermal-IR, the question of the orbital inclination can also impact the observations, although less severely than for coronagraphs thanks to the higher angular resolution. For inclined orbits the planet must be outside the instrument IWA ($\sim\lambda/B$, e.g., 20 mas assuming a typical (long) baseline $B \sim 100$ m at a wavelength of 10 μm). Thanks to formation flying, the interferometer baselines can be chosen in a broad interval, at least when $\theta_{HZ} \gg 20$ mas (see next section), so that the planet is seen under an angle significantly larger than the IWA during most of the time, e.g. 67% of the time for $\theta_{HZ} = 2$ x IWA. In case of a prior detection and identification of the relevant stellar systems, e.g., by an astrometric mission, the ephemerides will be known, allowing the selection of spectroscopic observations at dates when the systems are close to full elongation.

---

[3] based on the 380 F,G,K stars at D < 20 pc

Many nearby stars over-fulfilling the requirement $\theta_{HZ} > 20$ mas, *a detection by the spectroscopic mission itself could be possible*, by observing the targets three to four times to avoid unfavorable configurations. However, this would not be not optimum because we would ignore the mass of the planets (see Sect. 9).

The impact of the inclination of a planet orbit and the phase curve of the mid-IR signal with time of the planet, is not large when they it has an atmosphere. It is mainly that from a uniformly warm disc, because when not vanishing, the atmosphere redistributes the heat from the dayside to the nightside, and the thermal IR emission is similar when either side is observed. In a quantitative study of the Earth mid-IR spectrum as seen from distance, Gómez-Leal, Pallé, and Selsis (2012) found that spectra that change by few percent's with the spin rotation of our planet, mainly due to the type of landscape seen at a given time, e.g., continents/oceans, rather than the alternation of day and night. The IR emission comes from the upper parts of the atmosphere (the mid and lower parts are mostly opaque), whereas the day-night cycle mainly affects the boundary layer between atmosphere and ground (first kilometers) (same reference).

As for coronagraphs, a prior detection and orbital characterization of planets in the HZ of the target stars is assumed, and as much as possible, observations are performed close to the full elongation of the planet to allow a good angular separation of the planet from other sources.

## 6.3 Angular resolution of the star-planet system

The interferometer can reject the stellar light at the level $10^{-5}$ only if its short baseline, $B_1$, is not too large (Defrère et al. 2010), meanwhile, the long base, $B_2$, determines the angular resolution for resolving the planetary system. One of the advantage of the FF configuration is to allow a choice of these two quantities, independently. To avoid penalizing optical distortions, the maximum size of the long baseline that is compatible with the 1,200 m focal length considered in the *Darwin* project (Defrère et al. 2010) is $B_2 = 400$ m. Even in a version cheaper than the 2007 *Darwin* or TPF-I projects, with 0.75 m mirrors rather than $\geq 2$ m ones, the size of the interferometer geometry should be kept similar for preserving the resolution power of the instrument. This should be possible in the formation flying setup of the interferometer. Long baselines require good control of the attitude of the flotilla, therefore accurate metrologies, which would have some, but limited, impact on the cost, and a significant one on the performance. Accurate formation flying should be accessible in the near future thanks to the heritage of the PRISMA (Delpech et al. 2013) and PROBA-3 (ESA 2012) missions. Assuming a 4:1 ($B_2 : B_1$) configuration, which is understood as a minimum requirement to control instability noise (Lay 2006), the maximum baseline $B_2$ is 400 m. To be observed, the HZ must be seen by the interferometer as a few resolution elements ($\lambda/B_2$), 9 mas @ 18 μm, and 5 mas @ 10 μm. If requirement is:

$$\theta_{HZ} > 20 \qquad [\text{mas}], \tag{Eq.12}$$

the minimum number of resolution elements will be k = 4.0 @ 10 μm, and k = 2.2 @ 18 μm, possibly sufficient to separate the planet from the (mostly symmetrical) Exo-zodi.

This requirement is satisfied for most FGK stars with rank < 100 (Table 9), but it is of special importance for low luminosity M stars ($a_{HZ} = 1.30\ L_i^{1/2}$ AU) (Table 8). Only stars respecting this constraint are considered in our target list.

## 6.4 Calibrating the integration time (value of $d$)

In a way similar to coronagraphs, the constant $d$ of Eq.6 can be determined by fitting the predictions of the model to the values given in Defrère et al. 2010 that are obtained with the *DarwinSim* software (den Hartog 2005). For 2 m telescopes, 2.0 $R_{earth}$ planets, Defrère et al. 2010 find that 36 G stars can be studied in spectroscopy ($\lambda/\Delta\lambda = 20$), in 1.5 yr (Table 4 therein).

Comparing with the model (Table 7) yields

$$d = 1.5\ \text{yr}\ /\ 4.44 \times 10^4 = 3.4 \times 10^{-5}\ \text{yr}$$

**Table 7**

Integration times divided by *d* for G stars[a]

| Hipp. | Common name | Dble-st (as) | D [pc] | θ_HZ (mas) | V | Lbol (L_sun) | SpType | ti/d | star # | S_(ti/d) |
|-------|-------------|--------------|--------|------------|------|--------------|--------|---------|--------|-----------|
| 71683 | alf Cen A | 10[b] | 1.29 | 984 | -0.01 | 1.61 | G2.0V | 1.43e+00 | 1 | 1.43e+00 |
| 8102 | tau Ceti | 137 | 3.65 | 197 | 3.49 | 0.52 | G8.5V | 2.31e+01 | 2 | 2.45e+01 |
| 96100 | sig Draco | ___ | 5.75 | 115 | 4.67 | 0.44 | G9.0V | 1.08e+02 | 3 | 1.33e+02 |
| (…) | | | | | | | | | (…) | |
| 67927 | ___ | ___ | 11.4 | 272 | 2.68 | 9.6 | G0IV | 2.70e+03 | 35 | 4.17e+04 |
| 22263 | ___ | ___ | 13.28 | 76 | 5.49 | 1.01 | G3V | 2.72e+03 | **36** | **4.44e+04** |
| 79537 | ___ | ___ | 13.89 | 33 | 7.53 | 0.21 | G8/K0V | 2.98e+03 | 37 | 4.73e+04 |
| (…) | | | | | | | | | (…) | |

[a] conditions are $\Phi$ = 2 m, Rpl = 2.0, $\lambda/\Delta\lambda$ = 20

[b] in year 2030

For the interferometer, the accuracy of the model can also be estimated by applying it to different interferometer designs. Defrère et al. 2010 considered several of them, and the associated numbers of accessible stars can be compared with the model predictions. For planets with radius 1.0, 1.5, and 2.0 R_earth around M stars, during 0.3 yr and $\eta_{earth}$ = 100%, Defrère et al. 2010 find *24, 44, and 69 planets*, respectively. With the same conditions and *d* fixed to $3.4 \times 10^{-5}$ yr, the model finds *24, 47, and 70 planets,* respectively. As in the case of coronagraphs, *the similarity between the two series indicates a fair robustness of the model*.

## 6.5 Distinguishing planet from exozodiacal dust

For the interferometer, the presence of exozodiacal dust around a star can also hamper the direct detection and characterization of telluric planets in the thermal IR. However, nulling interferometers are less affected by exozodiacal dust than coronagraphs for two reasons: (1) their internal modulation eliminates the signal from centrally symmetric sources, leaving only their quantum noise (Mennesson et al. 2005, Defrère et al. 2010); and (2) their angular resolution is determined by their baselines, not by the diameters of the light collectors (individual mirrors), and can be high.

In an affordable version of a formation flying nulling interferometer (Sect. 6.6), the mirrors must be reduced, e.g., to ~ 0.75 m, whereas the baselines can remain large, e.g., 400 m (Sect. 6.3). This provides many resolution elements in the field of view (FOV), which is about the HZ size, e.g., $(12)^2$ = 144 in the case considered by Defrère et al. (2012, Fig.4 therein), who concluded that "Unlike the coronagraph, there is no significant detection (SNR>3) of false positives beyond 20 zodis".

For parity with coronagraphs, we will ignore the problem of exozodiacal dust, keeping in mind that *it is significantly less severe than for coronagraphs, the two instruments in an affordable version.*

## 6.6 Performances of an affordable nulling interferometer

What could be an affordable interferometer?  By rough analogy with the 2.4 m class coronagraph and starshade that are presently studied by NASA, a 4 x 0.75 m interferometer is considered as affordable.  It is pointed that this *does not include the technologic effort* needed before building the instrument, to rise it to the required Technical Readiness Level (TRL).

As discussed in Sect. 6.3, there is a major advantage in keeping the possibility of long baselines even for nearby targets where short baselines are expected to be sufficient: if an interesting system is found, it will be possible to qualify/falsify it by using a *higher spatial resolution,* to better separate the planetary signal from that of the exozodiacal cloud.

The IR luminosity of a planet with given radius and albedo, located in the HZ of its star, is an intrinsic property of the planet, independent of the parent star features.  It is that of a ~ 300 K body, with a given size.  As the thermal IR luminosity of an M2 star is ~ 5 times lower than that of a G2 star (Sect. 3.3), the planet/star contrast is more favorable for M than for G stars.  In addition, M stars are more numerous than FGK stars, per unit volume, so that if all stars were considered on the criterion of integration time alone (Eq. 6 and 7), M stars would fully dominate the target list.

Planets in the HZ of M stars are expected to have their spin and orbital rotations phase locked (Ulmschneider 2006), a situation very different from that of Earth regarding their habitability.  They would have a hemisphere continuously irradiated and the other continuously in the dark, which may be considered as worse or better for harboring life, but for sure different.  An M star surface should present an environment very different from that of our planet.  Therefore, it seems desirable to balance M and solar-types in the target list in order to study the diversity of different situations[4].  The two groups are considered separately, and it is proposed to spend the appropriate time on each of them to balance the numbers of stars in the two groups. Spending 4 yrs on FGK stars and 1 yr on M stars leads approximately to that result (Tables 8 and 9).

---

[4] The frontier between stars having, or not, phase locked planets in their HZ is complex.  For $a_{HZ}$ = 1.3 $L^{1/2}$  [AU], the frontier would be at the limit between M and K stars only in the absence of other planets that could prevent this locking by maintaining some eccentricity.  This frontier is used thereafter, being warned of the corresponding simplification.  The knowledge of individual planetary systems, and an estimate of possible resonances, would allow a better selection.

Table 8 illustrates the case of M star planets for an observation time of 1 yr, $R_{pl}$ = 1.5, the condition $\theta_{comp}$ > 18.5 - $\Delta m_{18}$ [as] but for Prox Cen, and $\theta_{HZ}$ > 20 [mas]. The impact of $\eta_{earth}$ is calculated as in Sect. 4.6 and shown for $\eta_{earth}$ = 100% and 10%, leading to 14 and 3.6 planets, respectively.

The same procedure is shown in Table 9 for FGK stars, and an observation time of 4 yrs, $R_{pl}$ = 1.5. For $\eta_{earth}$ = 100% and 10%, one gets $N_{pl}$ = 14 and 3.3, respectively.

**Table 8[a]**

M stars for 1 yr observations with a $\Phi$ = 4 x 0.75 m interferometer, $R_{pl}$ = 1.5 $R_{earth}$

| Hipparcos | Common name | D (pc) | V | L_bol (L_sun) | teta_HZ (mas) | SpType | Teff (K) | ti (yr) | (i) | S (ti) (yr) |
|---|---|---|---|---|---|---|---|---|---|---|
| 70890 | prox Cen | 1.30 | 11.01 | 8.56E-04 | 29.3 | M5.0V | 2425 | 1.25E-03 | 1 | 1.25E-03 |
| 87937 | Barnard's Star | 1.83 | 9.51 | 3.69E-03 | 43.2 | M3.5V | 2578 | 4.92E-03 | 2 | 6.17E-03 |
| 54035 | Lalande 21185 | 2.54 | 7.49 | 2.46E-02 | 80.3 | M2.0V | 3136 | 1.85E-02 | 3 | 2.47E-02 |
| 92403 | Ross 154 | 2.97 | 10.50 | 4.44E-03 | 29.2 | M3.5 V | 2614 | 3.41E-02 | 4 | 5.88E-02 |
| 114046 | Lacaille 9352 | 3.28 | 7.35 | 4.15E-02 | 80.7 | M1.0V | 3361 | 5.13E-02 | 5 | 1.1E-01 |
| 57548 | Ross 128 | 3.35 | 11.10 | 4.39E-03 | 25.7 | M4.0V | 2612 | 5.51E-02 | 6 | 1.65E-01 |
| 91772 | ___ | 3.52 | 9.70 | 9.02E-03 | 35.1 | M3.5V | 2790 | 6.73E-02 | 7 | 2.32E-01 |
| 91768 | ___ | 3.52 | 8.94 | 1.26E-02 | 41.5 | M3.0V | 2892 | 6.74E-02 | 8 | 3.00E-01 |
| 1475 | GX Andromedae | 3.57 | 8.13 | 2.38E-02 | 56.2 | M1.5V | 3122 | 7.15E-02 | 9 | 3.71E-01 |
| 36208 | Luyten's Star | 3.76 | 9.87 | 1.14E-02 | 36.9 | M3.5V | 2860 | 8.76E-02 | 10 | 4.59E-01 |
| 24186 | Kapteyn's Star | 3.91 | 8.85 | 1.20E-02 | 36.4 | M2.0V | 2876 | 1.02E-01 | 11 | 5.61E-01 |
| 110893 | Kruger 60 A | 4.03 | 9.59 | 1.42E-02 | 38.4 | M3.0V | 2931 | 1.16E-01 | 12 | 6.77E-01 |
| 30920 | Ross 614 A | 4.09 | 11.07 | 7.61E-03 | 27.7 | M4.5V | 2742 | 1.23E-01 | 13 | 8.00E-01 |
| 80824 | Wolf 1061 | 4.27 | 10.08 | 1.20E-02 | 33.4 | M3.5V | 2877 | 1.46E-01 | 14 | 9.46E-01 |
| 439 | ___ | 4.34 | 8.56 | 2.28E-02 | 45.3 | M1.5V | 3106 | 1.56E-01 | 15 | 1.10E+00 |
| 85523 | ___ | 4.54 | 9.41 | 1.70E-02 | 37.4 | M2.5V | 2996 | 1.86E-01 | 16 | 1.29E+00 |
| 86162 | ___ | 4.54 | 9.15 | 2.24E-02 | 42.8 | M3.0V | 3098 | 1.86E-01 | 17 | 1.47E+00 |
| 113020 | Ross 780 | 4.66 | 10.18 | 1.53E-02 | 34.5 | M2.5V | 2957 | 2.07E-01 | 18 | 1.68E+00 |
| 54211 | ___ | 4.86 | 8.82 | 2.28E-02 | 40.4 | M1.0V | 3105 | 2.45E-01 | 19 | 1.93E+00 |
| 106440 | ___ | 4.95 | 8.67 | 3.04E-02 | 45.8 | M1.5V | 3223 | 2.64E-01 | 20 | 2.19E+00 |
| 86214 | ___ | 5.05 | 10.94 | 1.04E-02 | 26.2 | M4.0 | 2831 | 2.85E-01 | 21 | 2.47E+00 |
| 112460 | EV Lacertae | 5.05 | 10.33 | 1.37E-02 | 30.1 | M3.5V | 2920 | 2.85E-01 | 22 | 2.76E+00 |
| 57544 | ___ | 5.34 | 10.80 | 8.40E-03 | 22.3 | M3.5V | 2769 | 3.56E-01 | 23 | 3.12E+00 |
| 67155 | Wolf 498 | 5.41 | 8.43 | 3.92E-02 | 47.6 | M1.0V | 3335 | 3.76E-01 | 24 | 3.49E+00 |
| 21088 | Stein 2051 | 5.54 | 10.82 | 1.12E-02 | 24.9 | M4.0V | 2856 | 4.12E-01 | 25 | 3.90E+00 |
| 33226 | ___ | 5.61 | 9.89 | 1.75E-02 | 30.7 | M3.0V | 3006 | 4.34E-01 | 26 | 4.34E+00 |
| 25878 | Wolf 1453 | 5.68 | 7.97 | 6.69E-02 | 59.2 | M1.0V | 3594 | 4.58E-01 | 27 | 4.80E+00 |
| 103039 | ___ | 5.71 | 11.46 | 7.68E-03 | 20.0 | M3.5V | 2745 | 4.65E-01 | 28 | 5.26E+00 |
| 29295 | ___ | 5.75 | 8.15 | 5.74E-02 | 54.2 | M1.5V | 3517 | 4.81E-01 | 29 | 5.74E+00 |
| 86990 | ___ | 5.84 | 10.78 | 9.54E-03 | 21.7 | M3.0V | 2806 | 5.09E-01 | 30 | 6.25E+00 |
| 94761 | Wolf 1055 | 5.85 | 9.12 | 3.37E-02 | 40.8 | M2.5V | 3267 | 5.14E-01 | 31 | 6.76E+00 |
| 73182 | ___ | 5.86 | 8.07 | 1.28E-01 | 79.4 | M1.5V | 3952 | 5.22E-01 | 32 | 7.29E+00 |
| 76074 | ___ | 5.93 | 9.31 | 3.16E-02 | 39.0 | M2.5V | 3239 | 5.42E-01 | 33 | 7.83E+00 |
| 117473 | ___ | 5.95 | 8.98 | 2.72E-02 | 36.0 | M1.0V | 3176 | 5.49E-01 | 34 | 8.38E+00 |
| 37766 | Ross 882 | 5.98 | 11.23 | 1.34E-02 | 25.2 | M4.0V | 2913 | 5.6E-01 | 35 | 8.94E+00 |
| 45343 | ___ | 6.11 | 7.64 | 8.20E-02 | 60.9 | M0.0V | 3701 | 6.14E-01 | 36 | 9.55E+00 |
| 34603 | QY Aurigae A | 6.12 | 11.67 | 1.21E-02 | 23.4 | M4.5V | 2880 | 6.14E-01 | 37 | 1.02E+01 |
| 99701 | ___ | 6.20 | 7.97 | 6.72E-02 | 54.3 | M0.0V | 3596 | 6.5E-01 | 38 | 1.08E+01 |
| 71253 | HN Librae | 6.22 | 11.32 | 1.16E-02 | 22.5 | M4.0V | 2866 | 6.55E-01 | 39 | 1.15E+01 |
| 74995 | Wolf 562 | 6.34 | 10.57 | 1.32E-02 | 23.6 | M3.0V | 2908 | 7.07E-01 | 40 | 1.22E+01 |

[a] a list up to rank 180 is available in the electronic version

**Table 9** [(a)]

FGK stars for 4 yr observations with a $\Phi$ = 4 x 0.75 m interferometer, $R_{pl}$ = 1.5 $R_{earth}$

| Hippar cos | Common name | Dbl-st (as) | d (pc) | HZ (mas) | V | L_bol (L_Sun) | SpType | star # | ti (yr) | S_(ti) (yr) |
|---|---|---|---|---|---|---|---|---|---|---|
| 71681 | alf Cen B | 10 (2030) | 1.29 | 562 | 1.35 | 0.52 | K0V | 1 | 2.11E-03 | 2.11E-03 |
| 71683 | alf Cen A | 10 (2030) | 1.29 | 984 | -0.01 | 1.61 | G2.0V | 2 | 3.21E-03 | 5.32E-03 |
| 104217 | 61 Cyg B | 31.6 | 3.5 | 89 | 5.95 | 0.1 | K7.0V | 3 | 6.72E-02 | 7.25E-02 |
| 104214 | 61 Cyg A | 31.6 | 3.5 | 105 | 5.2 | 0.13 | K5.0V | 4 | 6.77E-02 | 1.4E-01 |
| 108870 | eps Indi A | 400 | 3.62 | 131 | 4.69 | 0.23 | K4V | 5 | 7.96E-02 | 2.2E-01 |
| 8102 | tau Ceti | 137 | 3.65 | 197 | 3.49 | 0.52 | G8.5V | 6 | 8.69E-02 | 3.07E-01 |
| 105090 | AX Micro | ___ | 3.95 | 71 | 6.69 | 0.08 | K9.0V | 7 | 1.08E-01 | 4.15E-01 |
| 49908 | ___ | ___ | 4.87 | 67 | 6.6 | 0.11 | K7.0V | 8 | 2.49E-01 | 6.64E-01 |
| 19849 | omic 2 Eri | ___ | 4.98 | 131 | 4.43 | 0.42 | K0.5V | 9 | 2.85E-01 | 9.49E-01 |
| 96100 | sigma Dra | ___ | 5.75 | 115 | 4.67 | 0.44 | G9.0V | 10 | 5.00E-01 | 1.45E+00 |
| 73184 | ___ | 25.6 | 5.86 | 92 | 5.72 | 0.29 | K4.0V | 11 | 5.31E-01 | 1.98E+00 |
| 99461 | ___ | ___ | 6.01 | 88 | 5.32 | 0.28 | K2.5V | 12 | 5.88E-01 | 2.57E+00 |
| 120005 | ___ | ___ | 6.11 | 46 | 7.7 | 0.08 | K7.0V | 13 | 6.14E-01 | 3.18E+00 |
| 15510 | 82 Eri | ___ | 6.04 | 138 | 4.26 | 0.69 | G8.0V | 14 | 6.18E-01 | 3.8E+00 |
| 99240 | delta Pav | ___ | 6.11 | 189 | 3.53 | 1.33 | G8.0IV | 15 | 6.75E-01 | 4.47E+00 |
| 114622 | ___ | ___ | 6.54 | 84 | 5.57 | 0.31 | K3.0V | 16 | 8.21E-01 | 5.30E+00 |
| 12114 | ___ | ___ | 7.18 | 75 | 5.82 | 0.29 | K3V | 17 | 1.19E+00 | 6.49E+00 |
| 3765 | ___ | ___ | 7.45 | 73 | 5.74 | 0.3 | K1V | 18 | 1.38E+00 | 7.87E+00 |
| 7981 | ___ | ___ | 7.53 | 91 | 5.24 | 0.47 | K1V | 19 | 1.45E+00 | 9.32E+00 |
| 5336 | ___ | ___ | 7.55 | 90 | 5.17 | 0.46 | G5Vp | 20 | 1.46E+00 | 1.08E+01 |
| 113283 | ___ | ___ | 7.61 | 59 | 6.48 | 0.2 | K4VP | 21 | 1.49E+00 | 1.23E+01 |
| 88574 | ___ | ___ | 7.76 | 24 | 9.36 | 0.04 | K4/5V | 22 | 1.59E+00 | 1.39E+01 |
| 2021 | beta Hyd | ___ | 7.46 | 258 | 2.82 | 3.7 | G1IV | 23 | 1.61E+00 | 1.55E+01 |
| 5496 | ___ | ___ | 8.19 | 25 | 9.82 | 0.04 | K | 24 | 1.98E+00 | 1.74E+01 |
| 113576 | ___ | ___ | 8.22 | 38 | 7.87 | 0.1 | K5/M0V | 25 | 2.01E+00 | 1.95E+01 |
| 22449 | 1 Ori | ___ | 8.07 | 215 | 3.17 | 3 | F6V | 26 | 2.10E+00 | 2.16E+01 |
| 86974 | ___ | ___ | 8.31 | 200 | 3.41 | 2.78 | G5IV | 27 | 2.34E+00 | 2.39E+01 |
| 61317 | beta CVn | ___ | 8.44 | 133 | 4.24 | 1.27 | G0V | 28 | 2.34E+00 | 2.62E+01 |
| 113229 | ___ | ___ | 8.62 | 18 | 10.38 | 0.03 | K | 29 | 2.42E+00 | 2.87E+01 |
| 64924 | ___ | ___ | 8.56 | 109 | 4.74 | 0.87 | G5V | 30 | 2.43E+00 | 3.11E+01 |
| 1599 | ___ | ___ | 8.59 | 134 | 4.23 | 1.33 | G0V | 31 | 2.51E+00 | 3.36E+01 |
| 32984 | ___ | ___ | 8.71 | 56 | 6.58 | 0.24 | K3V | 32 | 2.55E+00 | 3.61E+01 |
| 23311 | ___ | ___ | 8.71 | 62 | 6.23 | 0.3 | K4III | 33 | 2.55E+00 | 3.87E+01 |
| 99825 | ___ | ___ | 8.91 | 73 | 5.72 | 0.42 | K3V | 34 | 2.81E+00 | 4.15E+01 |
| 47103 | ___ | ___ | 9.02 | 14 | 10.91 | 0.02 | K | 35 | 2.91E+00 | 4.44E+01 |
| 57939 | ___ | ___ | 9.09 | 53 | 6.42 | 0.23 | G8Vp | 36 | 3.02E+00 | 4.74E+01 |
| 27072 | ___ | ___ | 8.93 | 178 | 3.59 | 2.51 | F7V | 37 | 3.03E+00 | 5.05E+01 |
| 23512 | ___ | ___ | 9.21 | 11 | 11.73 | 0.01 | K: | 38 | 3.15E+00 | 5.36E+01 |
| 15457 | ___ | ___ | 9.14 | 102 | 4.84 | 0.88 | G5V | 39 | 3.15E+00 | 5.68E+01 |
| 64394 | ___ | ___ | 9.13 | 133 | 4.24 | 1.48 | G0V | 40 | 3.20E+00 | 6.00E+01 |

(a) a list up to rank 200 is available in electronic form

### 6.6.1 Assuming that $\eta_{earth}$ is common for M stars and F G K stars

Table 10 and Fig.6 give the number of planets that can be studied, as a function of a common $\eta_{earth}$ for FGK and M stars, for various planetary radii. The dependence of $N_{pl}$ on $\eta_{earth}$ can also be fitted by a power law with a $\beta_{interf} = 0.62$ exponent, indicating a somewhat slower dependence than for the coronagraph ($\beta_{coron} = 0.71$)

**Table 10**

Number of accessible planets for a common $\eta_{earth}$

| $\eta_{earth}$ (%) | $R_{pl} = 1.0$ | $R_{pl} = 1.5$ | $R_{pl} = 2.0$ |
|---|---|---|---|
| 1 | 0.49 // 0.52 | 0.96 // 0.88 | 1.6 // 1.2 |
| 3 | 0.93 // 0.96 | 1.83 // 1.83 | 2.9 // 2.7 |
| 10 | 2.1 // 2.0 | 3.3 // 3.6 | 6.0 // 6.0 |
| 30 | 4.2 // 3.6 | 7.5 // 7.2 | 11.2 // 11.1 |
| 100 | 10 // 7 | 14 // 14 | 24 // 24 |

**Note:** FGK stars are left // M stars are right

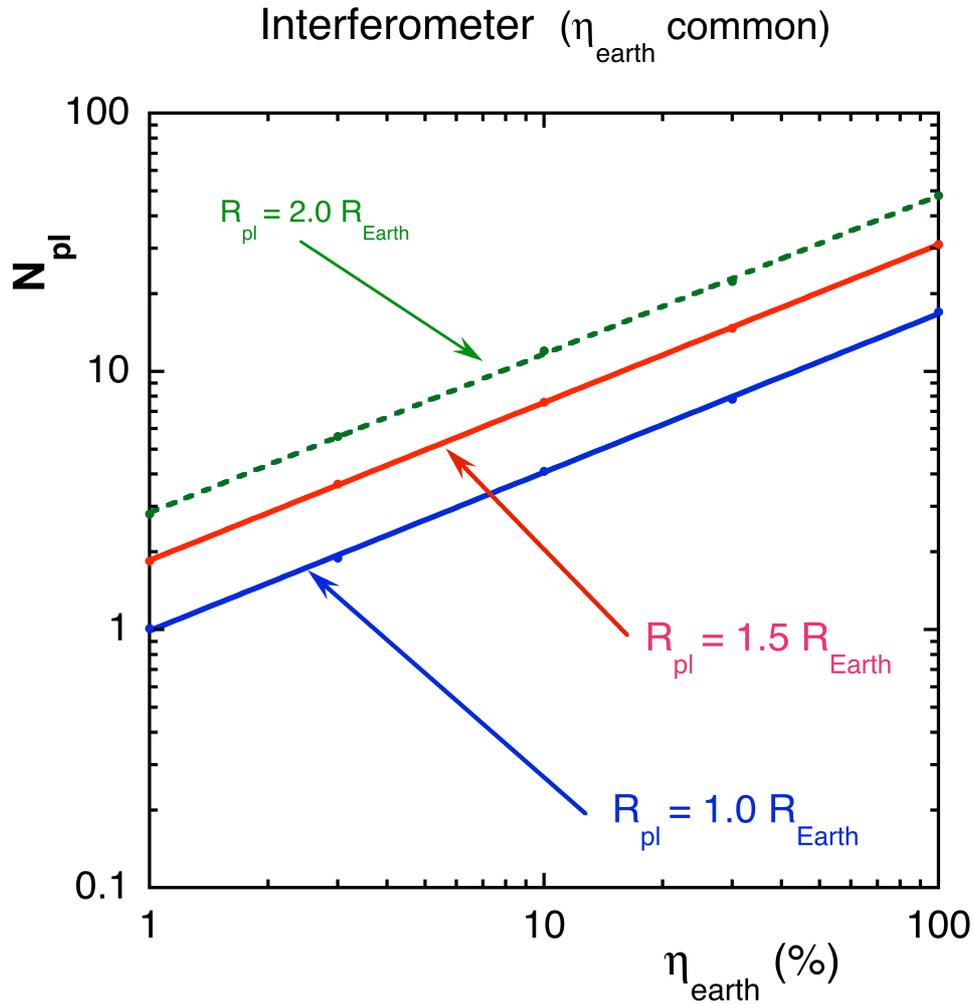

 Total number of planets located in the HZ of stars that can be studied in spectroscopy ($\lambda/\Delta\lambda = 20$), in the thermal IR, with a nulling interferometer having four 0.75 m mirrors, versus an $\eta_{earth}$ parameter that is common to all spectral types. 4 yrs are spent on FGK stars and 1 yr on M stars. As in Fig.4, the case of $R_{pl} = 2.0\ R_{Earth}$ planets is represented with a dashed line to indicate that they are probably not rocky (Rogers 2015, Fig.2 therein). Data points are calculated from the target list, and are well fitted by power laws with an $\beta_{inter} = 0.62$ exponent (lines).

### 6.6.2 Assuming that η$_{earth}$ is 50% for M stars, but unknown for F G K stars

In the present state of *Kepler* data analysis, $\eta_{earth}$ is better constrained for M stars than for FGK stars because the orbital period of planets in the HZ around the former is a few weeks rather than few months. Present Kepler's catalogue is considered as sufficient for addressing their statistics (Batalha 2014). Gaidos (2013) found $\eta_M \sim 50\%$. This is in agreement with estimate by RV technique, $\eta_M = 0.41_{+0.54/-0.13}$ (Bonfils et al. 2013). An alternative to the procedure of Sect.6.6.1 is to fix $\eta_M = 50\%$ for M stars, and leave $\eta_{solar}$ variable for FGK stars.

The corresponding number of planets is shown in Table 11 and Fig.7. As expected, the total number of planets is significantly larger for low values of $\eta_{solar}$ than in Sect. 6.6.1. For instance 13 planets can be studied for $\eta_{solar} = 10\%$ and $R_{pl} = 1.5$, instead of 7.1 for $\eta_M = \eta_{solar} = 10\%$.

**Table 11**

Case η$_M$ = 50%, open η$_{solar}$

| η$_M$ = 50%, | R$_{pl}$ = 1.0 | R$_{pl}$ = 1.5 | R$_{pl}$ = 2.0 |
|---|---|---|---|
| η$_{solar}$ [%] | | | |
| 1 | 0.5 // 4.5 | 1.0 // 10 | 1.6 // 15.5 |
| 3 | 0.9 // 4.5 | 1.8 // 10 | 2.9 // 15.5 |
| 10 | 2.1 // 4.5 | 3.3 // 10 | 6.0 // 15.5 |
| 30 | 4.2 // 4.5 | 7.5 // 10 | 11.2 // 15.5 |
| 100 | 10 // 4.5 | 14 // 10 | 24 // 15.5 |

**Note:** FGK stars are left // M stars are right

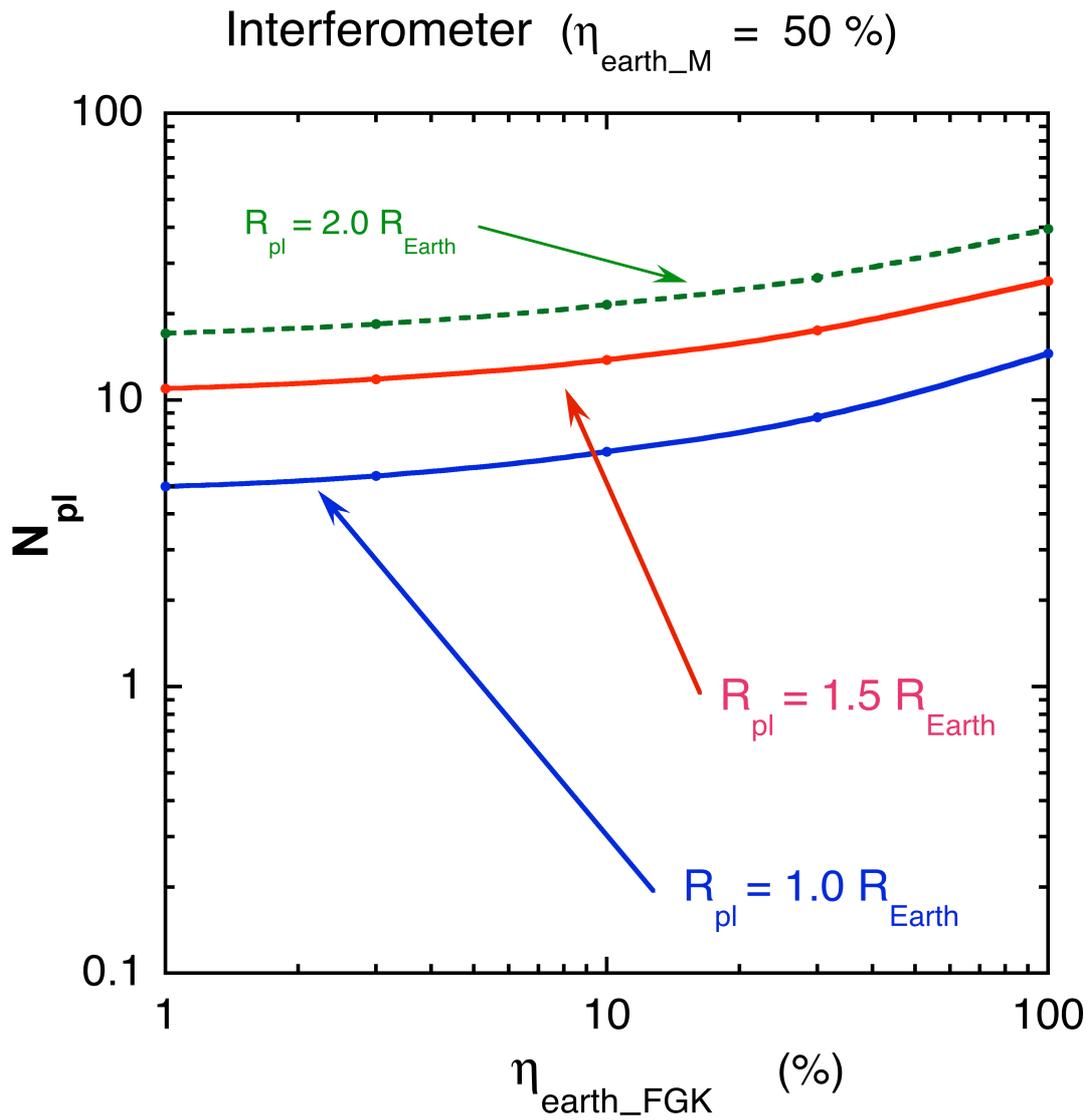

**6.7 Detections by the interferometer**

In a way similar to the case of coronagraphs, one can estimate how many stars could be investigated by the spectroscopic mission itself to search for planets, being aware of the associated disadvantages, *in particular the absence of knowledge of the planetary masses* (Sect. 9). Assuming that this could be done with three visits to avoid phase problems, and a spectral resolution $\lambda/\Delta\lambda = 2$, the proportionality of the required observation time to $\Delta\lambda$ (Eq. 6) and the need for three visits give the observing time from Table 8 and 9.

If 0.5 yr is dedicated to planet detection around M stars, *the 4 x 0.75 m interferometer could search for 1.5 $R_{earth}$ planets, up to the star ranked 19* (Ross 780, D = 2.97 pc), but for $\eta_M = 50\%$, only 5.5 planets could be studied in spectroscopy during the remaining 0.5 yr, instead of 7.5 if planets are identified beforehand.

If 2.0 yrs are dedicated to the search of planets around FGK stars, 17 stars could be surveyed (up to Hip12114, D = 7.18 pc), but for $\eta_{solar} = 10\%$ and $R_{pl} = 1.5$ $R_{earth}$, only 2.5 planets could be studied during the 2.0 yrs left for spectroscopy, instead of 3.3 if planets are identified beforehand.

Around all targets, M and FGK stars, 8 telluric planets could be studied in spectroscopy without prior detection, but it would be at the cost of ignoring their mass and therefore their physical nature.

# 7. Mission requirements for a given number of planets to be studied

Another interesting use of the model is the search for the requirements on an instrument with a given objective in terms of number of planets that can be studied. Only the case of $\eta_{solar} = 10\%$, $\eta_M = 50\%$, will be considered, with two instruments: a built-in coronagraph and an interferometer.

The procedure is as follows. Eq.5, or Eq.6, gives the integration time per target. New tables are built, analogous to Table 3, 8, and 9, but for various parameters including the mirror diameter. For a given goal of $N_{pl}$ planets, the requirements for cases with $\eta_{solar} \neq 1$ and planets identified beforehand, are obtained by studying $N_{pl}$ x $1/\eta_{earth}$ stars, during virtual sequences of observation $1/\eta_{earth}$ longer than the allocated time for that sequence (Sect. 4.6). The case of 1 yr spent on M stars and 4 yrs on FGK stars is considered, both for the coronagraph and the interferometer. The number of accessible planets is obtained by adjusting $N_{pl}$ so that the total durations are those allocated.

## 7.1 Coronagraphs

The impact of the mirror diameter on coronagraphs is twofold: (i) the IWA (2.5 $\lambda/\Phi$) decreases with increasing $\Phi$, which has an important impact on the transmission $\rho$*(D)* (Eq. 10), and (ii) the integration time varies according to Eq. 5 ($\propto D^{-2}$). The resulting number of accessible planets in spectroscopy is plotted in Fig.8, as a function of the mirror diameter.

*The goal of 10 planets* (7 around FGK stars, and 3 around M stars) *requires a mirror diameter of 5.5 m corresponding to a very expensive mission that is not likely to be affordable in the near future[5].* A goal of 3 planets (all around FGK stars) would still require a 3.5 m telescope.

---

[5] even if some enthusiasts are considering telescopes with diameters up to 10 m (Atlast 2009).

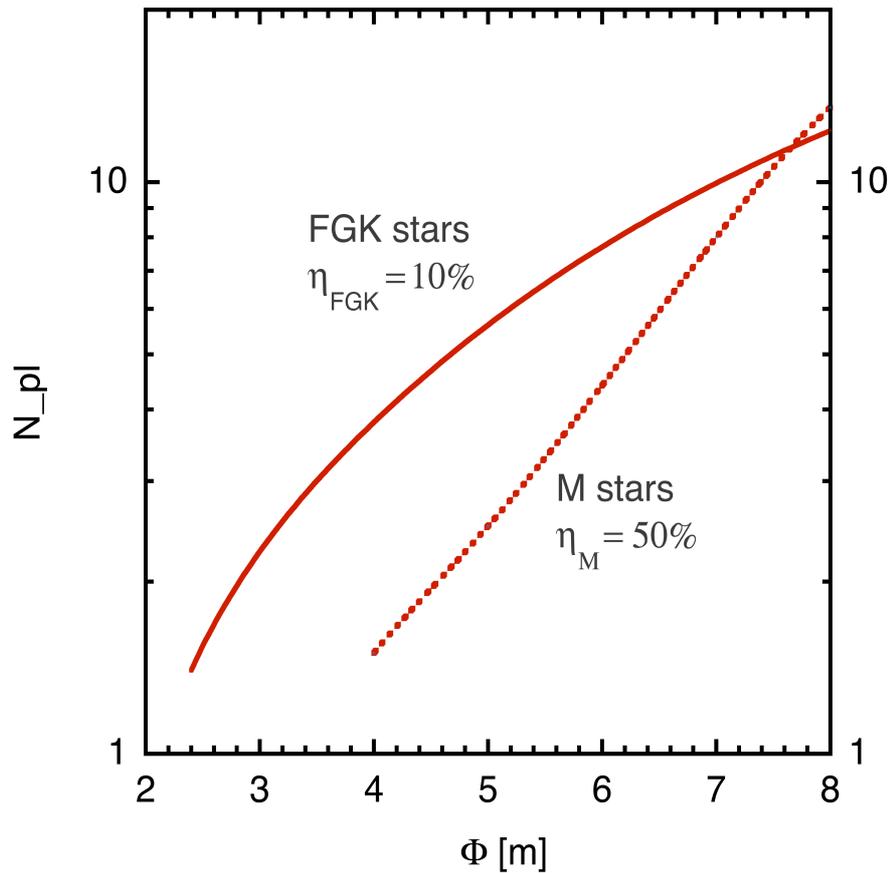

Figure 8: Number of planets around FGK and M stars that can be studied in spectroscopy by a coronagraph, as a function of the mirror diameter $\Phi$, for IWA = 2.5 $\lambda/\Phi$, $\lambda$ = 0.8 µm (O$_2$ spectral band A), resolution $\lambda/\Delta\lambda$ = 70, 1.5 R$_{earth}$ radius planets located at 1.3 L$^{1/2}$ [AU], a 5 yr mission, $\eta_{solar}$ = 10%, $\eta_M$ = 50%, and a prior identification of the suitable stars. M stars appear in the target list for diameter ≥ 4 m, and their fraction increases rapidly for larger diameters.

## 7.2 Comparison with another published model

The model can also be applied to the cases of coronagraphs recently considered by Stark et al. (2014), and its results compared with those of these authors. With the same conditions: detection only ($\lambda/\Delta\lambda$ = 5), $\lambda$ = 0.55 μm, IWA = 2.0 $\lambda/\Phi$, $\eta_{solar}$ = 10%, ignoring the need for several visits to avoid the planetary orbital phase problem, and searching for strict Earth analogues ($R_{pl}$ = 1.0, $a$ = 1.0 $L^{1/2}$ AU) during a 5 yr mission, the present model (Stark et al.'s model) finds that a 4 m coronagraph can discover 6.2 (4.0) planets, a 8 m coronagraph 18 (15) planets, and a 15 m coronagraph 48 (50) planets, respectively. *This reasonable agreement backs up both independent approaches.*

## 7.3 Interferometers

For interferometers, the mirror size impacts the quantum noise, but not the angular resolution when using formation flying (Sect.6.3). Eq. 6 determines the impact of $\Phi$ on the detection time in the target list, and Eq.7 gives the number of possible targets. For diameters smaller than 1.5 m, noise is dominated by the first term in the sum of Eq. 6 (the solar zodiacal emission, Defrère et al. 2010), so that an integration time $\propto \Phi^{-4}$ results. The number of planets that can be investigated in spectroscopy around M stars is calculated from Table 8, considering a longer virtual mission (Sect.4.6) of 1 x 2 x $(\Phi/0.75m)^4$ yr. For FGK stars Table 9 is used, with time: 4 x 10 x $(\Phi/0.75m)^4$ yr. The number of planets is the product of the target rank by $1/\eta_{earth}$. The cases of both M and FGK stars are shown in Fig.9.

The goal of *10 planets* (7.5 around M stars, 2.5 around FGK) requires *0.63 m mirrors*. With *1.0 m mirrors* the spectroscopic study of *21 planets*, could be done, 5.5 planets around FGK stars and 15.5 around M stars.

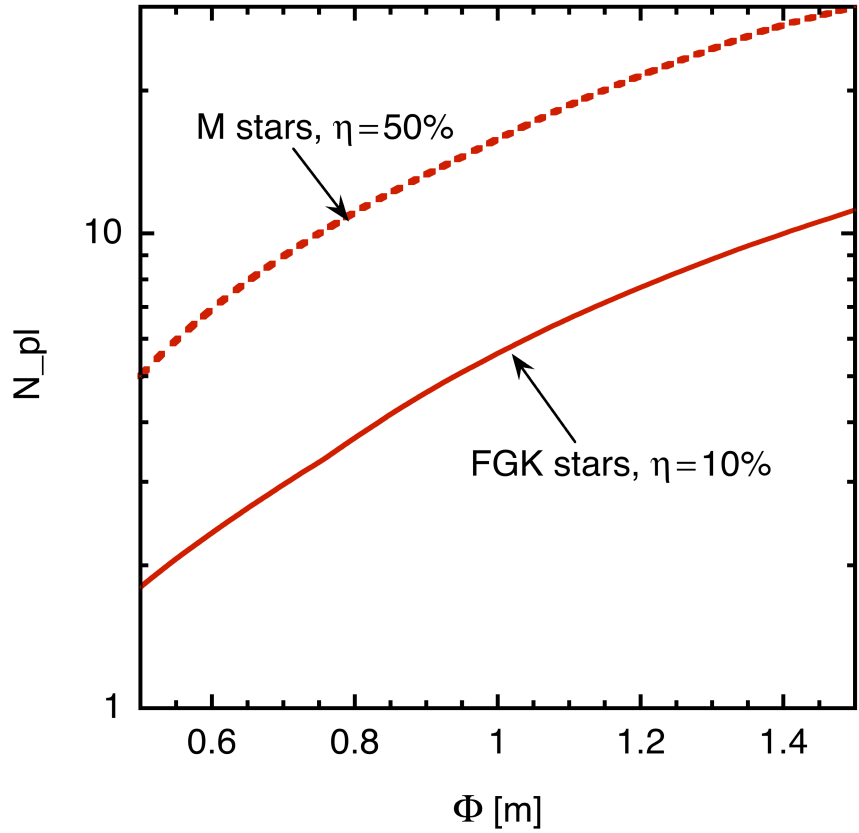

Figure 9: the number of planets that can be studied by an interferometer around FGK and M stars, as a function of the diameter of the four collecting mirrors, for a spectral resolution $\lambda/\Delta\lambda$ = 20 (spectroscopy), 1.5 $R_{earth}$ planets located at 1.3 $L^{1/2}$ [AU], and a prior identification of the suitable stars. The mission is 5 yr long, 1 yr is spent on M stars with $\eta_M$ = 50%, 4 yrs are spent on FGK stars with $\eta_{solar}$ = 10%.

# 8. Capabilities of Ground based instruments

The advent of extremely large telescopes on the ground in a foreseeable future, e.g., the 39 m E-ELT with first light expected in 2024 (http://www.eso.org/public/teles-instr/e-elt/), and its diffraction-limited resolution in the infrared, should also be considered for direct observation of small planets, specially in systems with a reduced star/planet contrast, as M stars. Quanz et al. (2014) state that METIS [the mid-infrared E-ELT imager and spectrograph] might be the first instrument to image a nearby (super-) Earth-sized planet with an equilibrium temperature near that expected to enable liquid water on a planet surface''.

This is hoped for imaging in broad bands, but ground-based, spatially resolved spectroscopy is *not expected to be applicable to the search for biosignature(s),* as the Earth atmosphere is fully opaque in most of the relevant spectral bands, such as $CO_2$ (15 $\mu$m) for instance. The atmospheric bands of the gas are optically thick and enlarged by collisions with other gaseous species at 1 bar pressure, preventing the detection of extra-terrestrial $CO_2$ through the comb of the atmospheric lines.

There is another niche for ground observations, the observation of planetary atmospheres during transits. The contrast is different because (i) the source is much brighter (the star, not the illuminated planet) but the cross-section is lower (the atmospheric layer, not the planetary disc), (ii) the effective altitude (pressure) for the line-widths is higher, 24 km, (lower, 65 mbar) for an Earth-like planet (Ehrenreich et al. 2006), and lines are narrower. However, the present situation is unclear. Based on the Earth-shine observations on the Moon, Arnold et al. (2014) concluded that a ground-based detection of the strongest oxygen $O_2$-A band seems possible with the E-ELT in the visible for an Earth-like planet in front of a G2 star at 10 pc, that of $O_3$ and $H_2O$ being challenging. On the other hand, Rodler & Lopez-Morales (2014) showed that $O_2$-A detection with an ELT could be feasible only for Earth-twins orbiting a close-by (D < 8 pc) M-dwarf with a spectral type later than M3.

We conclude that if ground-based ELTs could make the first direct image of an Earth-sister planet and possibly perform spectroscopy of $O_2$ in the visible for some transiting planets around (probably late M-type) nearby stars, they cannot replace space-based observatories for an exhaustive search of biosignatures by spectroscopy.

# 9. Discussion and conclusion

Our parametric model estimates the capabilities of instruments able to perform the spectroscopy of habitable planets, which could be affordable in the mid-future, e.g., coronagraphs and starshade in the visible with a $\Phi$ = 2.4 m mirror, and nulling interferometers in the thermal-IR with 4 x 0.75 m telescopes. Their goal is searching for biosignatures in the atmosphere of terrestrial planets located in the HZ of nearby stars, and we study the expected mission yield as a function of $\eta_{earth}$, and of the radius of planets.

The selected target list is specific to each instrument. Starting from the list of the nearest stars, the fist step is to select the sub-sample that can be observed with each instrument. For both types of coronagraphs, there is no M star, because the HZ around these stars is seen under too small an angle, around 20 to 60 mas for the nearest M stars (Table 8), whereas the IWA is 170 mas for the built-in coronagraph with $\Phi$ = 2.4 m, IWA = 2.5 $\lambda/\Phi$, and $\lambda$ = 0.8 μm; or 115 mas for a 34 m starshade. For the interferometer, the target list comprises both M stars and solar-type (FGK) stars due to adjustable baseline length.

Mission yields depend on $\eta_{earth}$. For large values, say $\eta_{earth}$ = 50%, all three instruments can study a reasonable number of planets, around 5, 10, and 20, respectively, *assuming that the problem of separating the planet from the exozodiacal emission can be solved for each of them*. This would justify a dedicated mission in all cases.

For low value of $\eta_{earth}$, coronagraphs and interferometers are not equally equipped. For $\eta_{earth}$ = 10%, *both built-in and starshade coronagraphs with 2.4 m mirrors can study only a very limited sample, 1.5 and 2.0 planets respectively,* for $R_{pl}$ = 1.50 $R_{earth}$. This would probably be prohibitive for dedicated missions, but a mission with other astrophysical goals could be worth it.

For the presently favored values by Kepler, $\eta_M$ = 50% for M stars, and $\eta_{solar}$ = 10% for FGK stars, an interferometer with 4 x 0.75 m mirrors *could study ~ 14 planets*. Even with the risks of small number statistics, several objects of very high scientific significance could be studied, which *would legitimate a dedicated mission.*

The capabilities of both types of instrument are summarised in Table 12 for these values of $\eta_{earth}$; and in the Annex 2, Table A1, for $\eta_M$ = 50% and $\eta_{solar}$ = 20%.

**Table 12**

Mirror sizes for different goals, in planet number $N_{pl}$.

| Goal<br><br><br>($N_{pl}$) | Coronagraph:<br>required<br>$\Phi_{coro}$<br>(m) | Composition<br>(S) = solar star<br>(M) = M star | Interferometer:<br>required<br>$\Phi_{inter}$ (x 4)<br>(m) | Composition<br>(S) = solar star<br>(M) = M star |
|---|---|---|---|---|
| 1 | 2.2 | (S): 1.0<br>(M): 0.0 | 0.13 | (S): 0.4<br>(M): 0.6 |
| 3 | 3.1 | (S): 2.4<br>(M): 0.6 | 0.17 | (S): 1.0<br>(M): 2.0 |
| 10 | 5.5 | (S): 6.6<br>(M): 3.4 | 0.62 | (S): 2.5<br>(M): 7.5 |
| 30 | 8.4 | (S): 13<br>(M): 17 | 1.20 | (S): 8.0<br>(M): 22 |

**Note:** for $R_{pl}$ = 1.5 $R_{earth}$, $\eta_M$ = 50%, and $\eta_{solar}$ = 10%.

However, the maturity of the technology required for the different missions is not the same. Thanks to large efforts, built-in coronography is closer to maturity. Relevant contrasts have been obtained in the laboratory with instruments that are not far from being space qualified. The WFIRST-AFTA project (WFIRST 2014) should fly with such a coronagraph but the central obstruction, will make the $10^{-10}$ rejection not feasible. The starshade technology is presently being developed at JPL (NASA 2014).

For the interferometer, the maturity is lower. At JPL, Martin et al. (2012) demonstrated in the laboratory the capacity of cancelling the stellar light and recovering that of an artificial planet, with the contrast needed in the thermal IR ($\sim 10^{-7}$). However, this was made at high fluxes, e.g., using $CO_2$ lasers. To go further and demonstrate that capacity with astronomical fluxes, a cryogenic experiment is needed, which is not planed. In addition, the instrument must be compact and space qualified. There is no identified show-stopper, but an important technology effort is required to reach the appropriate TRL.

On the other hand, *formation flying* is not that far from maturity. In 2010, the success of the Swedish mission PRISMA qualified the navigation at the centimetre level using radio-frequency sensors for inter-satellite distances ranging from 2 m to 10,000 m, without loss nor collision. When the distance was maintained fixed, the standard deviation in X, Y, and Z was ~ 2.5 cm for satellites in a low Earth orbit (LEO), which is not a gravitationally quite environment. During a series of simulated astronomical observations, the pointing of the two satellites was set in predefined directions with a typical accuracy of 1.7 arc-minute (Delpech et al. 2013). In 2017, the approved PROBA-3 mission (ESA 2012) should qualify it at the level of laser sensors (0.1 mm), which is more than enough for the requirements of nulling interferometers having internal delay-lines.

Most of the estimates made in this paper are made under the hypothesis that *a prior mission has detected the planets of interest*.

Building *both* types of spectroscopic instruments, *coronagraph and interferometer*, would be, by far, the best solution because they are complementary (Beichman et al. 2007). Most of the scientific community agrees that this is what we need in the long future. The two instruments work in different parts of the electromagnetic spectrum, and each one is better suited for detecting a given group of gases, $O_2$, $H_2O$, $CH_4$ for coronagraphs, $O_3$, $CH_4$, $CO_2$, for interferometers ($H_2O$ can be detected only with difficulties). Having access to both wavelength ranges on the same planets would provide a much firmer basis for our conclusions. Fortunately, there is some overlap between the two lists of accessible solar-type stars. For instance, among the 15 best stars in the built-in coronagraph list, (Table 3), 10 are also in the interferometer list limited to rank 100 (full Table 9).

Now, the present paper addresses the question: which affordable first mission can provide the largest harvest? Would $\eta_{earth}$ be confirmed small for FGK stars, a built-in coronagraph on a large (~ 5 m) mirror, or a $\geq$ 60 m starshade, would be necessary to obtain a significant sample of planets (~10), whereas a middle-sized interferometer, e.g., 4 x 0.6 m, could do it (Table 12).

This also stresses the interest of getting to a consensus on the value of $\eta_{earth}$ from *Kepler* data. The values of $\eta_{solar}$ already obtained by Petigura et al. (2013) and Silburt et al. (2015) are in good agreement when the same definition of the HZ is adopted. If the definition used is that of Sect.2: $a_{in} = 0.99\ L_i^{1/2}$, $a_{out} = 1.70\ L_i^{1/2}$ for a Sun-like spectrum and $R_{pl} = 1.0 - 2.0\ R_{earth}$ planets, these authors find $\eta_{solar} = 8.6\% \pm 3\%$, and $\eta_{solar} = 6.4\%$ +3/-1%, respectively. Were these numbers the ultimate values, Fig.4 and Fig.7 would indicate the number of planets that could be studied by each instrument.

In summary, Table 12 *points out that the interferometer remains an important part of the (Darwin, TPF-I) / (TPF-C, Starshade) debate.*

## Acknowledgements

The authors would like to sincerely thank an anonymous referee whose comments led to substantial improvements in the manuscript, Luc Arnold for discussions on the E-ELT capabilities, Franck Selsis for discussions on the thermal-IR emission of planets with atmosphere during their day-night cycle, and Antoine Crouzier for different comments. A.L. and F.M. thank CNES for their continuous technical support regarding space interferometry, free flying, and high precision astrometry. O.A. acknowledges support from the European Research Council under the European Union's Seventh Framework Program (ERC Grant Agreement n. 337569) and from the French Community of Belgium through an ARC grant for Concerted Research Action.

## Supplement information

In the electronic version, full Tables 3, 8, and 9 are available in machine readable form, with target stars ranked by increasing integration time.

- Full_table_3:  coronagraph, 2.4 m mirror, 100 FGK stars

- Full_table_8:  interferometer, 4 x 0.75 m mirrors, 180 M stars

- Full_table_9:  interferometer, 4 x 0.75 m mirrors, 200 FGK stars

## Annex 1: Minimum distance of a stellar companion for a built-in coronagraph

The number of photo-electrons on the detector from the planet (signal), from the target star, and from a companion star are, in number:

$$N_{pl} = (1/128\,h\nu)\,A\,Y\,L'_* \,(R_{pl}\,/\,a)^2 (\Phi\,/\,D)^2\,t$$

$$N_* = (1/16\,h\nu)\,\rho\,Y\,L'_* \,(\Phi\,/\,D)^2\,t$$

$$N_{comp} = (1/16\,h\nu)\,\text{Airy}(\theta)\,Y\,L'_{comp}\,(\Phi\,/\,D)^2\,t$$

where $L'$ are the stellar luminosities within the spectral band of the instrument (0.6 – 0.8 μm), approximately the R band for Si-based detectors, and Airy(θ) the Airy function.

- Using $F_0 = 3.6\ 10^{-9}$ [W m$^{-2}$] for a R = 0 object, the luminosity of a R magnitude star at distance $D$ [m] reads:

$$L' = 4.5\,10^{-8}\ 10^{-0.4R}\ D^2 \quad [\text{W}]$$

- Photo-electrons from the planet are:

$$N_{pl} = 3.0\ 10^8\ (D\,/\,1\,\text{pc})^{-2}\ (\Phi\,/\,1\,m)^2\ (t\,/\,1\text{d}) \quad [\text{ph-el}]$$

- From a companion, at angular distance θ, they are

$$N_{comp} = 1.19\,10^{17}\ \text{Airy}(\theta)\ 10^{-0.4\,R_{comp}}\ (\Phi\,/\,1\,\text{m})^2\ (t\,/\,1\text{d})$$

Requiring a signal/noise, $N_{pl}/(N_{comp})^{1/2}$, of 10, the Airy function must decrease the companion flux by a factor:

$$\text{Airy}(\theta) = 1.83\,10^{-8}\ (D\,/\,1\,\text{pc})^{-4}\ 10^{0.4\,R_{comp}}\ (\Phi\,/\,2.4\ \text{m})^2\ (t\,/\,1\text{d})$$

The Airy envelop at large angle θ is: Airy(x) = (8/π) x$^{-3}$, x ≡ θ Φ/λ.  With the reference values for the nearest target stars: $D = 3$ pc, and $R_{comp} = 3$, and a 2.4 m mirror, the angular separation of a companion must satisfy Eq.8:

$$\theta_{comp} > 35.5\ (D\,/\,3\ \text{pc})^{1.33}\ 10^{-0.13\,(R_{comp}-3)}\ (\Phi\,/\,2.4\ \text{m})^2\ (t\,/\,1\text{d})^{-1/3} \quad [\text{as}]$$

**Annex 2**

**Table A1**

Same as Table 12, but for $\eta_{solar}$ = 20%

| Goal<br><br>($N_{pl}$) | Coronagraph:<br>required<br>$\Phi_{coro}$<br>(m) | | Composition<br>(S) = solar star<br>(M) = M star | Interferometer:<br>required<br>$\Phi_{inter}$ (x 4)<br>(m) | | Composition<br>(S) = solar star<br>(M) = M star |
|---|---|---|---|---|---|---|
| 1 | | 1.6 | (S): 1.0<br>(M): 0 | 0.15 | | (S): 0.5<br>(M): 0.5 |
| 3 | | 2.5 | (S): 2.5<br>(M): 0 | 0.25 | | (S): 1.5<br>(M): 1.5 |
| 10 | | 4.2 | (S): 8<br>(M): 2 | 0.55 | | (S): 4<br>(M): 6 |
| 30 | | 7.2 | (S): 21<br>(M): 9 | 1.05 | | (S): 13<br>(M): 17 |